\documentclass[a4paper,fleqn]{cas-sc}

\usepackage[numbers]{natbib}

\usepackage{amssymb}
\usepackage{amsmath}
\usepackage{upgreek}
\usepackage{algorithmicx}
\usepackage{algorithm}
\usepackage{algpseudocode}
\usepackage{hyperref}
\usepackage{graphicx}
\usepackage{booktabs} 
\usepackage{xcolor} 

\DeclareMathOperator*{\argmin}{arg\,min}

\def\tsc#1{\csdef{#1}{\textsc{\lowercase{#1}}\xspace}}
\tsc{WGM}
\tsc{QE}

\begin{document}
\let\WriteBookmarks\relax
\def\floatpagepagefraction{1}
\def\textpagefraction{.001}

\shorttitle{Self-Adaptive Model-Order Reduction for Shock-Dominated Problems}    

\shortauthors{}  

\title [mode = title]{Self-Adaptive Model-Order Reduction for Shock-Dominated Problems}  



%
    \author[1]{Ali Mohaghegh}[
      orcid=0009-0009-9321-5134,
]


\cormark[1]


\ead{ali.mohaghegh@ku.edu}



\affiliation[1]{organization={University of Kansas},
            addressline={Aerospace Engineering}, 
            city={Lawrence},
            postcode={66045}, 
            state={KS},
            country={USA}}

    \author[2]{Cheng Huang}[
      orcid=0000-0000-0000-0000,
]


\ead{chenghuang@ku.edu}


\credit{}

\affiliation[2]{organization={University of Kansas},
            addressline={Aerospace Engineering}, 
            city={Lawrence},
            postcode={66045}, 
            state={KS},
            country={USA}}

\cortext[1]{Corresponding author: Ali Mohaghegh}



\begin{abstract}
Projection-based model-order reduction (MOR) offers a promising approach for accelerating expensive full-order model (FOM) by constructing reduced-order models (ROMs) that retain the majority of FOM accuracy while significantly reducing the computational cost. However, conventional static ROMs, often following offline/online training strategy, can exhibit limited predictive capability for convection-dominated problems featuring strongly evolving physics such as shocks. Adaptive ROMs emerge as a promising alternative and address such limitation by dynamically updating the low-dimensional subspace during online simulation to track locally evolving system dynamics, thereby enabling accurate predictions even for shock-dominated flows. However, such online adaptation introduces additional computational overhead, and frequent basis and sampling-point updates can substantially reduce their computational advantage relative to their static counterpart. To reconcile this tradeoff, the current work introduces a \textit{self-adaptive MOR framework} that dynamically constructs and deploys the most appropriate ROM formulation (\emph{either} adaptive \emph{or} static) \emph{on the fly} according to the evolving characteristics of the solution. to achieve optimal computational efficiency while maintaining high modeling accuracy of the local physics during the online model execution. In this regard, the proposed framework seeks to exploit the advantages of both adaptive- and static-ROM approaches to achieve an optimal computational efficiency and predictive accuracy while reducing reliance on extensive offline training. The framework is developed and evaluated for modeling rotating detonation engines (RDEs), which provide a challenging test environment involving strongly convection-dominated, transient, and multiscale reacting-flow dynamics.
\end{abstract}



\begin{keywords}
 \sep Computational Fluid Dynamics (CFD) \sep Model-Order Reduction \sep Reduced-Order Modeling \sep Machine Learning \sep Shock-Dominated Flows \sep Rotating Detonation Engine \sep Turbulent Combustion
\end{keywords}

\maketitle

\section{Introduction}\label{sec:intro}

Rapid advances in high-performance computing technologies have enabled high-fidelity simulations of increasingly complex engineering systems \cite{oefelein_advances_2019,aditya_direct_2019,wang_towards_2017}. However, the direct integration of these simulations in many-query applications, such as design optimization and uncertainty quantification, remains computationally prohibitive for complex engineering systems because repeated model evaluations are required over a wide range of parameters and geometric configurations to examine key distinct system dynamics. In this regard, projection-based model order reduction (MOR) has emerged as a promising systematic approach to reduce this computational burden by constructing efficient reduced order models (ROMs) from full-order models (FOMs) through rigorous mathematical model reduction \cite{lumley_low-dimensional_1997,benner_survey_2015,mohaghegh_machine_2025}. As demonstrated by numerous studies in literature \cite{benner_survey_2015,mcquarrie_data-driven_2021,carlberg_galerkin_2017}, the resulting ROMs retain majority of the modeling fidelity of FOM while achieving orders-of-magnitude acceleration in computational efficiency. The construction of projection-based ROMs typically follows an offline/online strategy: (1) during the \emph{offline} stage, a limited number of high-fidelity FOM simulations are performed over a target parameter space to generate training snapshots, which are used to construct a global low-dimensional subspace, commonly through proper orthogonal decomposition (POD) \cite{sirovich_turbulence_1987,berkooz_proper_1993} or greedy reduced-basis methods \cite{prudhomme_reliable_2002, veroy_certified_2005,binev_convergence_2011}; (2) during the \emph{online} stage, the governing equations of FOM are projected onto the precomputed low-dimensional subspace, which remains \emph{static} and is used to construct the efficient ROM. Though this strategy proves to be highly effective for problems possessing rapidly decaying low-dimensional structures (e.g., diffusion-dominated dynamics), it presents three major challenges for complex convection-dominated problems.



\emph{First}, the commonly used linear subspace exhibits poor performance in representing convection-dominated dynamics, especially these featuring sharp gradients (e.g., flames) and discontinuities (e.g., shocks). Consequently, the required basis dimension can become sufficiently large that the computational advantages of model reduction and the associated hyper-reduction for nonlinear problems diminish \cite{barrault_empirical_2004, chaturantabut_nonlinear_2010}. This behavior is often referred as the \textbf{Kolmogorov barrier} \cite{peherstorfer_breaking_2022,greif_decay_2019}, which has become a crucial bottleneck for modern ROM development since the convection-dominated dynamics play more important roles as the systems become increasingly complex, multi-physics and multi-scale \cite{cohen_optimal_2020,bonnaillie-noel_efficient_2016}. This so-called Kolmogorov barrier has been observed across a wide range of application domains, including reacting and strongly convective flows \cite{huang_exploration_2018, huang_predictive_2023}, wildfire dynamics \cite{black_efficient_2021}, pulsed detonation combustors \cite{king_model_2019}, rotating detonation waves \cite{king_model_2019, mendible_dimensionality_2020, camacho_investigations_2024, mohaghegh_self_2026}, kinetic plasma problems \cite{nicolini_reduced-order_2021}, space-weather prediction \cite{issan_predicting_2023} and pattern formation \cite{geelen_localized_2022}.


\emph{Second}, constructing a sufficiently representative low-dimensional subspace can itself incur \textbf{prohibitively high training cost}. Specifically, the accuracy of the ROM heavily relies on whether the \emph{offline} training snapshot ensemble adequately samples 
the dynamics encountered \emph{online}. For accurate modeling of parametric behaviors in multi-physics and multi-scale systems, the creation of such subspace may require long trajectories, fine temporal 
sampling, and repeated FOM evaluations across the target
parameter space. Yet these are precisely the systems for which each individual high-fidelity simulation is most expensive and becomes a major computational undertaking by itself. For example, 
a representative large eddy simulation (LES) of a rotating detonation engine (RDE) requires approximately one million CPU hours to simulate one millisecond of 
physical time using 14,336 CPU cores on a high-performance computing system, which takes about three days of wall-clock 
time per run \cite{farcas_distributed_2025}. Therefore, generating tens or hundreds of 
such trajectories to cover a parameter space of interest becomes
infeasible, \emph{while} in reality, due to limitations on data storage, each available individual training dataset often has to be 
temporally down-sampled and only confined to a narrow slice of the operating envelope 
\cite{farcas_parametric_2023}. This creates a fundamental dilemma in the classical 
offline/online strategy for projection-based MOR: the problems that benefit most from model reduction are 
often those for which generating a sufficiently rich training dataset is 
least affordable. 

\emph{Third}, conventional static-basis ROMs often present \textbf{limited predictive capabilities} to represent dynamics absent or deviated from the training dataset. This limitation becomes particularly distinct for complex systems when the parameter space is too large to be sampled exhaustively, when only a small number of trajectories can be generated using FOM, or when the online trajectory evolves beyond and deviates from the regime represented by the offline training snapshots \cite{peherstorfer_online_2015, peherstorfer_model_2020,amsallem_interpolation_2008, washabaugh_nonlinear_2012, qian_certified_2017}. Specifically, for multi-physics and multi-scale systems featuring convection-dominated physics, this limitation can lead to rapid deterioration of long-horizon predictive accuracy \cite{huang_predictive_2023,camacho_investigations_2024}. Improving the predictive capability of a static-basis ROM generally requires enriching the offline training dataset to encompass a broader range of possible dynamics. This requirement, however, further increases the cost of training-data generation and therefore reinforces the fundamental dilemma identified in the second challenge.
    



Several remedies have been proposed to address the aforementioned challenges in constructing effective ROM for convection-dominated problems. \emph{One direction} seeks to replace the commonly used linear subspace with advanced \textbf{nonlinear manifolds} with the goal of obtaining more accurate approximation of the target high-dimensional dynamics. To date, a variety of methods have been proposed and proved effective in constructing such nonlinear manifolds by (1) explicitly incorporating quadratic nonlinearities \cite{rutzmoser_generalization_2017, barnett_quadratic_2022,geelen_operator_2023}, (2) directly using convolutional autoencoders \cite{lee_model_2020,kim_fast_2022,fresca_pod-dl-rom_2022,buchfink_model_2024}, and (3) leveraging artificial neural networks (ANNs) to more efficiently parameterize low-dimensional subspaces \cite{barnett_neural-network-augmented_2023,halder_reduced-order_2024}. These methods can substantially improve low-dimensional representation of nonlinear and convection-dominated dynamics and can potentially enable further reduction in ROM dimensionality, while maintaining a high level of accuracy. However, the nonlinear parametrizations required by these type of methods can introduce additional challenges in constructing computationally efficient and robust ROM for complex systems. For example, quadratic manifolds require the computation and storage of additional matrices whose size scales quadratically with the reduced dimension, potentially leading to substantial memory footprint for large-scale problems \cite{aretz_nested_2026,geelen_operator_2023}. Similarly, a small latent dimension obtained using autoencoders does not necessarily translate into online efficiency because nonlinear lifting operations generally cannot be assembled from precomputed quantities \cite{lee_model_2020}. This lifting bottleneck often necessitates development of specialized hyper-reduction techniques, such as masked or compressed decoders and over-collocation \cite{kim_fast_2022,romor_non-linear_2023}. Moreover, since the nonlinear manifold is not a vector space, standard projection operator and orthogonality properties developed for linear-subspace ROMs do not directly apply, which makes stability analysis and rigorous error certification more challenging for these nonlinear-manifold ROMs \cite{black_projection-based_2020, klein_multi-fidelity_2026}. Most importantly, though the nonlinear-manifold ROMs exhibit improved compression and representation of available training dataset, their construction generally follow the offline/online strategy and therefore their performance still heavily reply on the representativeness of the offline training dataset. Consequently, using nonlinear manifold for ROM construction alone does not fundamentally resolve the challenges associated with limited training data or predicting dynamics outside the training regime.

\emph{One alternative direction} focuses on \textbf{adaptive MOR}, in which the low-dimensional subspace is updated dynamically online as new evolving dynamics are encountered \cite{mohaghegh_feature-guided_2026, singh_lookahead_2023, brand_fast_2006,sapsis_dynamically_2009,iollo_advection_2014,peherstorfer_online_2015,zucatti_adaptive_2023,huang_predictive_2023}. Unlike conventional static-basis ROMs, adaptive approaches do not require a fixed subspace to represent all possible dynamics \emph{a priori}; instead, they progressively enrich or modify the low-rank representation using information acquired during the online simulation. Early works explored hybrid FOM/ROM strategies in which an on-the-fly criterion determines when the FOM should be invoked to enrich an inadequate low-rank representation. For example, Bai et al. \cite{bai_deim_2020,feng_bai_reduced_2022} developed switching criteria based on rapid low-rank singular value decomposition (SVD) updates. Feng et al. \cite{feng_fomrom_2021} introduced a rigorous a \emph{posteriori} error estimator to systematically control this FOM/ROM switching mechanism. Yano et al. \cite{yano_globally_2021} developed a trust-region framework that guides the ROM construction process to satisfy designated accuracy requirements in topology optimization. Carlberg \cite{carlberg_adaptive_2015} presented an online basis enrichment strategy in which a reduced-basis vector is partitioned into multiple vectors with disjoint support, analogous to mesh-based h-refinement, which has been subsequently enhanced by Etter and Carlberg \cite{etter_online_2020} through a vector-space sieving procedure. 

For large-scale nonlinear systems, an important class of adaptive MOR methods seeks to update the subspace using only \emph{sparse information} from the FOM, thereby circumventing repeated evaluation of the complete high-dimensional system. Peherstorfer and Willcox \cite{peherstorfer_online_2015} introduced the adaptive discrete empirical interpolation method (ADEIM), which extends the empirical interpolation method (EIM) \cite{barrault_empirical_2004,chaturantabut_nonlinear_2010} by inferring low-rank corrections to the reduced space from sparsely sampled residual information. Subsequent developments have addressed several key aspects of this online adaptation process, including adaptive selection of sampling locations \cite{peherstorfer_model_2020}, orthonormality-preserving low-rank basis updates using subspace-tracking techniques \cite{balzano_online_2010,zimmermann_geometric_2018}, and lookahead strategies that anticipate evolving dynamics to determine more effectively when and where FOM information should be sampled \cite{zucatti_adaptive_2023,singh_lookahead_2023}. More recent approaches have further specialized online adaptation for challenging convection-dominated systems containing moving sharp gradients, shocks, and detonation waves \cite{huang_predictive_2023,mohaghegh_feature-guided_2026}, while history-aware basis updates based on incremental SVD have improved both predictive accuracy and computational acceleration \cite{hedayat_history-aware_2026}. These developments have also been extended through in-context learning strategies that exploit self-generated trajectories during online prediction \cite{hedayat_-context_2026}.

In parallel, time-dependent basis approaches have been proposed in which low-rank representations are evolved directly from the governing equations. For example, Ramezanian et al. \cite{ramezanian_--fly_2021}, Naderi and Babaee \cite{naderi_adaptive_2023}, and Aitzhan et al. \cite{aitzhan_reduced_2025} derived and integrated evolution equations for time-dependent low-rank representations, sharing close connections with dynamical low-rank approximation methods. In addition, inspired by dynamic low-rank formulations, structure-preserving adaptive reduced-basis techniques have been developed to retain important properties of the underlying dynamical system during basis adaptation \cite{hesthaven_structure-preserving_2021,hesthaven_rank-adaptive_2022}. Subsequently, recent work by Pagliantini and Vismara \cite{pagliantini_fully_2025} demonstrates that such structures can also be preserved when adaptation is combined with sparse sampling strategies such as EIM. 

Recently, the concept of adaptivity has also begun to emerge in non-intrusive ROM formulations. Hedayat et al. \cite{hedayat_toward_2026} formulated a practical framework for constructing adaptive non-intrusive ROMs that periodically incorporates newly observed high-fidelity states through a moving data window. Their formulations adapt both the reduced subspace and the learned latent dynamics, either through sequential SVD based basis updates and operator-inference refitting or through joint Riemannian optimization of the encoder, decoder, and latent dynamics. Geelen and Willcox \cite{geelen_localized_2022} developed a localized operator-inference approach in which the reduced operators are learned offline over different regions of the state space and an appropriate local low-rank representation is selected adaptively tailoring to the evolving system dynamics. More recently, Aretz and Willcox \cite{aretz_nested_2026} exploited the nested operator inference method, which leverages the hierarchical structure in the reduced space to facilitate efficient offline updates and warm-started refinements of learned operators.

Collectively, adaptive MOR provides a promising pathway to address the fundamental limitations of classical static-basis ROMs. By acquiring information from the FOM and updating the low-rank representation as the simulation evolves, the adaptive ROM methods can substantially reduce \cite{huang_predictive_2023,mohaghegh_feature-guided_2026,hedayat_history-aware_2026} or in some formulations entirely eliminate \cite{ramezanian_--fly_2021} the dependence on an extensive offline training stage. More importantly, online adaptation enables the ROM to naturally incorporate previously unseen dynamics rather than requiring those dynamics to be represented \emph{a priori} by the offline training dataset, which inherently enhance predictive robustness for complex convection-dominated problems \cite{huang_predictive_2023,mohaghegh_feature-guided_2026}. However, this enhanced predictive capability in ROM introduces an important trade-off: \textbf{adaptation itself carries a non-negligible online computational cost}. Frequent acquisition of FOM information, adaptive sampling, and repeated basis updates can substantially reduce the acceleration achieved relative to conventional static-basis ROMs. Moreover, continuously performing these operations may be unnecessary when the system revisits dynamics that have already been encountered and adequately represented by the evolving reduced space. Existing approaches therefore expose a fundamental trade-off: \emph{static ROMs are computationally efficient but limited by their precomputed representation, whereas adaptive ROMs are predictive but incur persistent online adaptation costs}.


The present work addresses this trade-off by introducing a \textbf{self-adaptive MOR framework} that selectively activates online adaptation according to the dynamics encountered during the simulation. Instead of treating the ROM as either permanetly static or continously adaptive, the proposed framework dynamically transitions between the two operating modes. Specifically, when previously unseen dynamics are encountered, the adaptive ROM formulation is adopted to update the low-rank representation and acquire the necessary state information \emph{on the fly}. Once the evolving dynamics become sufficiently represented by the already acquired information, the framework constructs and transitions to a static (basis) ROM without incurring continued adaptation costs. Adaptation is reactivated when the existing representation becomes inadequate. The proposed framework seeks to largely retain the predictive capability of adaptive ROM while recovering the computational efficiency of static ROM when adaptation is not necessary. This selective adaptation mechanism is particularly advantageous for systems containing \emph{recurrent dynamics}. Rotating detonation engines (RDEs) provide a representative example: after an initial transient phase, detonation waves repeatedly propagate around an annular domain and exhibit dynamically similar spatial structures. In such scenario, a conventional static ROM must represent these translating structures globally and may therefore require a large basis, whereas a continuously adaptive ROM repeatedly incurs the cost of basis adaptation even after the recurring dynamics have already been learned. The proposed self-adaptive MOR framework instead uses adaptation to learn the evolving dynamics during the initial cycles and subsequently exploits recurrence by switching to the substantially less expensive static ROM. Thus, information acquired through adaptation becomes reusable knowledge rather than a transient correction applied only at the current time step. The primary innovation of this work is therefore not a new basis-update algorithm itself, but a framework for \emph{selectively deploying adaptation and systematically reusing dynamically acquired information}. Specifically, the proposed methodology: (1) identifies when the current low-rank representation is sufficient and adaptation can be suspended; (2) transitions between adaptive and static ROM formulations during a single simulation; and (3) exploits recurrent dynamics to reuse previously learned reduced representations, thereby reducing the cumulative cost of online adaptation while preserving predictive accuracy. The resulting framework provides a pathway toward ROMs that can learn previously unseen dynamics when necessary and efficiently exploit that knowledge when those dynamics recur.

The remainder of this paper is organized as follows. Sections \ref{sec:FOM}, \ref{sec:SROM} and \ref{sec:AROM} review the formulation of the FOM, static and adaptive ROMs respectively. Section \ref{sec:motiv_SAROM} introduces a motivating example for the self-adaptive MOR and section \ref{sec:SAROM} presents its formulation and computational algorithm. Section \ref{sec:results} presents numerical results of test cases with assessment of their robustness, accuracy and efficiency. Section \ref{sec:conclusion} provides concluding remarks and perspectives on future work.

\section{Full-order model}\label{sec:FOM}

We begin by introducing the full-order model (FOM) as a generic nonlinear time-dependent dynamic system to represent the original partial differential equation (PDE) describing the conservation laws (e.g., the Navier–Stokes equations),
\begin{equation}
\frac{\text{d}\mathbf{q}}{\text{d}t} = \mathbf{f}(\mathbf{q}, t) \quad \text{with} \quad \mathbf{q}(t = 0) = \mathbf{q}_0 ,
\label{eq:fom}
\end{equation}

\noindent where \( t \in [t_0, t_f] \) is the solution time of interest, and \( t_0, t_f \in \mathbb{R}_{\geq 0} \) are the initial and final solution time instances, respectively. \( \mathbf{q} \in \mathbb{R}^N \) represents the conservative state vector, with \( N \) as the total number of degrees of freedom. The vector \( \mathbf{q}_0 \in \mathbb{R}^N \) specifies the initial condition at \( t = 0 \). The nonlinear operator \( \mathbf{f}: \mathbb{R}^N \to \mathbb{R}^N \) encapsulates the spatial discretization of fluxes, source terms, and body forces in the conservation equations, as well as contributions from boundary conditions. The total number of degrees of freedom is factored as \( N = N_{var} \times N_{elem} \), where \( N_{elem} \) is the number of points in the spatial discretization of the original PDE, and \( N_{var} \) is the number of solution variables at each spatial location (e.g., density, momentum, total energy, and species mass fraction). 

To numerically integrate Eq. \ref{eq:fom} in time, various time-stepping schemes can be employed, such as implicit or explicit multistep methods or Runge–Kutta (RK) methods. In this work, for one-dimensional test cases, we adopt the explicit four-stage third degree strong stability preserving Runge-Kutta (four-stage SSP-RK3) method \cite{durran_numerical_2010}, a general form of which can be expressed as
\begin{equation} 
\mathbf{q}^{n} = \mathbf{q}^{n-1} + \Delta t \sum_{i=1}^{s} \beta_i \mathbf{f}(\mathbf{q}^{(i)}, t^{(i)}),
\label{eq:SSPRK_grouped}
\end{equation}

\noindent where the intermediate stage solutions \( \mathbf{q}^{(i)} \) are computed via
\begin{equation}
\label{eq:SSPRK_intem}
\mathbf{q}^{(i)} = \mathbf{q}^{n-1} + \Delta t \sum_{j=1}^{i-1} \alpha_{ij} \mathbf{f}(\mathbf{q}^{(j)}, t^{(j)}),
\end{equation}

\noindent and the coefficients \( \alpha_{ij} \) and \( \beta_i \) are prescribed by the chosen time-integration scheme \cite{durran_numerical_2010}. Here, \( \Delta t \) denotes the physical time step used in the numerical integration. In addition, for two-dimensional test problems, we adopt the implicit linear multi-step method, the detailed mathematical expression of which can be found in \cite{huang_model_2022}, where the
$l$-step can be expressed as
\begin{equation}
\mathbf{q}^{n}
+
\sum_{j=1}^{l} \gamma_j \mathbf{q}^{n-j}
=
\Delta t \delta_0 \mathbf{f}\left(\mathbf{q}^{n},t^{n}\right)
+
\Delta t
\sum_{j=1}^{l}
\beta_j
\mathbf{f}\left(\mathbf{q}^{n-j},t^{n-j}\right).
\label{eq:lmm}
\end{equation}

\noindent Coefficients $\gamma_j,\delta_j \in \mathbb{R}$ are determined based on $l$. The FOM equation residual
$\mathbf{r}:\mathbb{R}^{N}\rightarrow\mathbb{R}^{N}$ is defined as
\begin{equation}
\mathbf{r}\left(\mathbf{q}^{n}\right)
\triangleq
\mathbf{q}^{n}
+
\sum_{j=1}^{l}
\gamma_j \mathbf{q}^{n-j}
-
\Delta t \delta_0
\mathbf{f}\left(\mathbf{q}^{n},t^{n}\right)
-
\Delta t
\sum_{j=1}^{l}
\delta_j
\mathbf{f}\left(\mathbf{q}^{n-j},t^{n-j}\right).
\label{eq:fom_residual}
\end{equation}

\noindent The state variables, $\mathbf{q}^{n}$, are solved for at each time step so that
$\mathbf{r}\left(\mathbf{q}^{n}\right)=0$.

\section{Static projection-based model-order reduction (MOR)}\label{sec:SROM}

This section reviews the \emph{static} projection-based MOR formulations employed withn the proposed self-adaptive MOR framework. Specifically, we consider three types of low-dimensional subspaces for static ROM construction, including the conventional linear subspaces, quadratic nonlinear manifolds, and nonlinear error closure modeling in the latent space. Moreover, we emphasize that the primary objective of the current work is to develop a general framework for ROM construction and deployment rather than focusing on a particular static ROM formulation. Accordingly, the proposed framework is designed to accommodate different types of static ROMs, allowing the selection of the low-dimensional subspace to be tailored to the problem type, underlying dynamics, and computational scale.

\subsection{Projection-based MOR with linear low-dimensional subspace} \label{sec:linearPOD}

First, we consider obtaining a low-rank approximation of the full state ($\mathbf{q}$ from Eq. \ref{eq:fom}) using the conventional linear subspace or basis ($\mathbf{V} \in \mathbb{R}^{N \times n_p}$)
\begin{equation}
\mathbf{\tilde{q}}(t) = \mathbf{q}_{\text{ref}} + \mathbf{P}^{-1}\mathbf{V{q}}_r(t),
\label{std. main rom}
\end{equation}

\noindent where $\mathbf{{\tilde{q}}} \in \mathbb{R}^{N}$ represents the approximation of $\mathbf{q}$. The reference state, $\mathbf{q_{\text{ref}}}$, is a constant translation vector commonly used in fluid dynamics problems, often chosen as the initial solution, $\mathbf{q_{\text{ref}}} = \mathbf{q}(t = t_0)$, or the time-averaged solution defined by $\mathbf{q_{\text{ref}}} = \frac{1}{\Delta T} \int_{t_0}^{t_0 + \Delta T} \mathbf{q}(t) \, dt$, and $\mathbf{q}_r \in \mathbb{R}^{n_p}$ represents the reduced state where $n_p$ denotes the number of basis modes included (or equivalently the dimension of ROM). In the current work, the linear subspace or trial basis ($\mathbf{V}$) is computed using proper orthogonal decomposition (POD) \cite{berkooz_proper_1993,lumley_low-dimensional_1997} via singular value decomposition (SVD)
\begin{equation}
\label{eq:svd}
\argmin_{V \in \mathbb{R}^{N \times n_p}} \| \mathbf{PQ} - \mathbf{VV}^T \mathbf{PQ} \|_F \quad \text{s.t.} \quad \mathbf{V}^T \mathbf{V} = I,
\end{equation}

\noindent where the data snapshot matrix $\mathbf{Q}$ is defined such that each column corresponds to a solution snapshot $\mathbf{q'} = \mathbf{q}(t) - \mathbf{q}_\text{ref} \in \mathbb{R}^{N}$ at $N_T$ time instants 
\begin{equation}
\mathbf{Q} = [\mathbf{q'}(t_0), \dots,\mathbf{q'}(t_f)] \in \mathbb{R}^{N \times N_T}.
\end{equation}

\noindent Moreover, a scaling matrix $\mathbf{P}$ is applied to align the solution variables within $\mathbf{q'}$ so they maintain uniform magnitudes across different physical quantities in $\mathbf{Q}$, preventing discrepancy due to variables with inherently larger magnitudes, such as total energy. In this work, all quantities are normalized by their $L^2$ norms following \cite{lumley_low-dimensional_1997}
\begin{equation}
\mathbf{P} = \text{diag}(\mathbf{P}_1, \dots, \mathbf{P}_i, \dots, \mathbf{P}_{N_{\text{elem}}}),
\end{equation}

\noindent where each $\mathbf{P}_i = \text{diag}(\phi_{1, \text{norm}}^{-1}, \dots, \phi_{N_{\text{var}}, \text{norm}}^{-1})$. The $v$-th state variable $\phi_{v, \text{norm}}$ is described as follows
\begin{equation}
\phi_{v, \text{norm}} = \frac{1}{\Delta T} \int_{t_0}^{t_0 + \Delta T} \frac{1}{\Omega} \int_{\Omega} \phi'_v(x, t)^2 \, dx \, dt.
\end{equation}

The procedure for model reduction via Galerkin projection is presented for the continuous-time formulation of the full-order model given in Eq. \ref{eq:fom}. The approach starts by premultiplying Eq. \ref{eq:fom} with the scaling matrix $\mathbf{P}$ and then projecting the result onto the test space \(\mathbf{V}\)
\begin{equation}
\label{prj.rom}
\mathbf{V}^T \mathbf{P} \frac{\text{d} \mathbf{q}}{\text{d}t} = \mathbf{V}^T \mathbf{P} \mathbf{f}(\mathbf{q}, t).
\end{equation}

\noindent Applying the scaling matrix in Eq. \ref{eq:fom} is essential to appropriately weight the individual equations within the projected reduced system. Omitting this step can cause the reduced model to be dominated by equations with inherently larger magnitudes, such as the energy equation, potentially leading to substantial floating-point errors. Introducing the low-rank representation from Eq. \ref{std. main rom} into Eq. \ref{prj.rom} yields the following reduced-order ODE system
\begin{equation}
\label{eq:reduced_ode}
\frac{\text{d} {\mathbf{q}_r}}{\text{d}t} = \mathbf{V}^T \mathbf{P} \mathbf{f}(\tilde{\mathbf{q}}, t), \quad {\mathbf{q}_r}(0) = \mathbf{V}^T \mathbf{q}_0,
\end{equation}

\noindent where the dimension of the reduced ODE system, $n_p$, is much smaller than the dimension $N$ of the full-order model in Eq. \ref{eq:fom} (i.e., $n_p \ll N$).

\subsection{Projection-based MOR with quadratic nonlinear manifold} \label{sec:nonlinearQuadratic}

Second, we seek to construct a more accurate low-rank approximation of the full state using nonlinear manifold. Specifically, following the work by Geelen et al.~\cite{geelen_operator_2023}, we introduce an additional quadratic nonlinear structure in Eq.~\ref{std. main rom} to augment the accuracy of the low-rank approximation
\begin{equation}
\mathbf{\tilde{q}}(t) = \mathbf{q}_{\text{ref}} + \mathbf{P}^{-1}\big[\mathbf{V}\mathbf{q}_r(t)+\mathbf{\bar{V}}(\mathbf{q}_r(t) \otimes \mathbf{q}_r(t))\big],
\label{eq:quad rom}
\end{equation}
\noindent where $\otimes$ is the Kronecker product restricted to unique terms $\mathbf{\bar{V}}\in\mathbb{R}^{N \times m}$ represents a quadratic mapping operator with $m = n_p \times (n_p+1) / 2$, which is inferred by solving the following optimization problem,
\begin{equation}
\mathbf{\bar{V}}=\argmin_{\mathbf{\bar{V}}\in\mathbb{R}^{N\times m}}{\big\|\mathbf{Q} - \mathbf{q}_{\text{ref}} - \mathbf{P}^{-1}[\mathbf{V}\mathbf{Q}_r+\mathbf{\bar{V}}(\mathbf{Q}_r \otimes \mathbf{Q}_r )] \big\|_2^2},
\label{eq:quad rom min problem}
\end{equation}
\noindent where $\mathbf{Q}_r\in\mathbb{R}^{n_p \times N_T}=\mathbf{V}^T\mathbf{Q}$ is the projection of snapshot matrix ($\mathbf{Q}$) onto the reduced basis. In addition, as proved by Geelen et al.~\cite{geelen_operator_2023}, the orthogonality condition holds between the leading linear basis ($\mathbf{V}$) and the quadratic mapping operator ($\mathbf{\bar{V}}$) -- i.e., $\mathbf{V}^T \mathbf{\bar{V}} = \mathbf{0}$. Furthermore, as discussed by Barnett and Farhat~\cite{barnett_quadratic_2022} and Geelen et al.~\cite{geelen_operator_2023}, directly solving for $\mathbf{\bar{V}}$ may lead to overfitting issue during the offline training stage, which necessitates the introduction of regularization in the least-squares problem posed in Eq. \ref{eq:quad rom min problem}
\begin{equation}
\mathbf{\bar{V}}=\argmin_{\mathbf{\bar{V}}\in\mathbb{R}^{N\times m}}{\big\|\mathbf{Q} - \mathbf{q}_{\text{ref}} - \mathbf{P}^{-1}[\mathbf{V}\mathbf{Q}_r+\mathbf{\bar{V}}(\mathbf{Q}_r \otimes \mathbf{Q}_r )] \big\|_2^2}+\lambda \|\mathbf{\bar{V}}\|_2^2,
\label{eq:quad rom min problem reg}
\end{equation}
\noindent where $\lambda$ is a scalar regularization coefficient. Further details on the computation of $\mathbf{\bar{V}}$ can be found in Geelen et al.~\cite{geelen_operator_2023}. The projection of the governing equations onto the quadratic manifold follows the same Galerkin procedure as the linear ROM. Differentiating Eq.~\ref{eq:quad rom} with respect to time gives
\begin{equation}
\frac{\mathrm{d}\tilde{\mathbf{q}}}{\mathrm{d}t}
=
\mathbf{P}^{-1}
\left[
\mathbf{V}\frac{\mathrm{d}\mathbf{q}_r}{\mathrm{d}t}
+
\mathbf{\bar{V}}
\frac{\mathrm{d}}{\mathrm{d}t}
\left(
\mathbf{q}_r\otimes\mathbf{q}_r
\right)
\right].
\label{eq:quad_time_derivative}
\end{equation}

\noindent Substituting this expression into the governing equations and projecting onto the linear basis $\mathbf{V}$ yields
\begin{equation}
\mathbf{V}^{T}\mathbf{V}
\frac{\mathrm{d}\mathbf{q}_r}{\mathrm{d}t}
+
\mathbf{V}^{T}\mathbf{\bar{V}}
\frac{\mathrm{d}}{\mathrm{d}t}
\left(
\mathbf{q}_r\otimes\mathbf{q}_r
\right)
=\mathbf{V}^{T}\mathbf{P}
\mathbf{f}(\tilde{\mathbf{q}},t).
\end{equation}

\noindent Since the basis matrices are constructed such that
\begin{equation}
\mathbf{V}^{T}\mathbf{V}=\mathbf{I},
\qquad
\mathbf{V}^{T}\mathbf{\bar{V}}=\mathbf{0},
\end{equation}

\noindent the contribution associated with the time derivative of the quadratic correction vanishes during projection. The resulting reduced system is therefore
\begin{equation}
\frac{\mathrm{d}\mathbf{q}_r}{\mathrm{d}t} = \mathbf{V}^{T}\mathbf{P}
\mathbf{f}(\tilde{\mathbf{q}},t),
\qquad
\tilde{\mathbf{q}}
=
\mathbf{q}_{\mathrm{ref}}
+
\mathbf{P}^{-1}
\left[
\mathbf{V}\mathbf{q}_r
+
\mathbf{\bar{V}}
\left(
\mathbf{q}_r\otimes\mathbf{q}_r
\right)
\right].
\label{eq:quad_reduced_ode}
\end{equation}

\noindent Thus, the quadratic manifold retains the same reduced coordinates and the same form of the reduced time derivative as the conventional linear Galerkin ROM. Its benefit enters through the reconstruction of $\tilde{\mathbf{q}}$ with the additional quadratic term provides a more expressive representation of the full state vector, which can improve the accuracy of the nonlinear term $\mathbf{f}(\tilde{\mathbf{q}},t)$ on the right-hand side without introducing additional independent reduced variables.

\subsection{Projection-based MOR with nonlinear error closure modeling in the latent space} \label{sec:GPR_RBF}

Third, we seek to augment the low-rank approximation using linear POD subspace (i.e., Eq. \ref{std. main rom}) by modeling closure errors in the latent space following the work by Ares De Parga et al. \cite{ares_de_parga_nonlinear_2026}
\begin{equation}
\mathbf{\tilde{q}}(t) = \mathbf{q}_{\text{ref}} + \mathbf{P}^{-1}[\mathbf{V}\mathbf{q}_r(t)+\mathbf{\hat{V}}\mathcal{N}(\mathbf{q}_r(t))],
\label{eq:NL rom}
\end{equation}
\noindent where $\mathbf{V}\in\mathbb{R}^{N\times n_p}$ and $\mathbf{\hat{V}}\in\mathbb{R}^{N\times \hat{n}_p}$ represent the leading basis matrix and the residual basis matrix, respectively. The combination of these two matrices constitutes the SVD-derived basis $\mathbf{V}_{tot}\in\mathbb{R}^{N \times n_{tot}}$ -- i.e., $\mathbf{V}_{tot} = [\mathbf{V} \quad \mathbf{\hat{V}}]$, such that $n_p \ll \hat{n}_p \ll N$ and $n_p + \hat{n}_p = n_{tot}$. More importantly, $\mathbf{V}$ and $\mathbf{\hat{V}}$ satisfy the following orthogonality conditions:
\begin{equation}
\mathbf{V}^T \mathbf{V} = \mathbf{I}_{n_p}, \quad \mathbf{\hat{V}}^T \mathbf{\hat{V}} = \mathbf{I}_{\hat{n}_p}, \quad \text{and} \quad \mathbf{V}^T \mathbf{\hat{V}} = \mathbf{0},
\label{eq:NL rom ortho}
\end{equation}

\noindent where $\mathbf{I}_{n_p}$ and $\mathbf{I}_{\hat{n}_p}$ denote the identify matrices of dimension $n_p$ and $\hat{n}_p$, respectively. In Eq. \ref{eq:NL rom}, $\mathcal{N}(\mathbf{q}_r):\mathbb{R}^{n_p} \to \mathbb{R}^{\hat{n}_p}$ denotes a nonlinear map describing the relationship between the generalized coordinates ($\mathbf{\hat{q}}_r(t)\in\mathbb{R}^{\hat{n}_p}$) and their affine counterparts ($\mathbf{q}_r(t)\in\mathbb{R}^{n_p}$) such that $\mathbf{q}(t) = \mathbf{q}_{\text{ref}} + \mathbf{P}^{-1}[ \mathbf{V}\mathbf{q}_r(t) + \mathbf{\hat{V}} \mathbf{\hat{q}}_r(t) ]$ and $ \mathbf{\hat{q}}_r(t) \approx \mathcal{N}(\mathbf{q}_r)$. The projection of the these nonlinear error closure models proceeds in the same procedure as the quadratic manifold above. Differentiating Eq.~\ref{eq:NL rom} with respect to time gives
\begin{equation}
\frac{\mathrm{d}\tilde{\mathbf{q}}}{\mathrm{d}t} = \mathbf{P}^{-1}
\left[
\mathbf{V}\frac{\mathrm{d}\mathbf{q}_r}{\mathrm{d}t}
+
\mathbf{\hat{V}}
\frac{\mathrm{d}}{\mathrm{d}t}
\mathcal{N}(\mathbf{q}_r)
\right].
\label{eq:NL_time_derivative}
\end{equation}

\noindent Substitution into the governing equations followed by projection onto $\mathbf{V}$ gives

\begin{equation}
\mathbf{V}^{T}\mathbf{V}
\frac{\mathrm{d}\mathbf{q}_r}{\mathrm{d}t}
+
\mathbf{V}^{T}\mathbf{\hat{V}}
\frac{\mathrm{d}}{\mathrm{d}t}
\mathcal{N}(\mathbf{q}_r) = \mathbf{V}^{T}\mathbf{P}
\mathbf{f}(\tilde{\mathbf{q}},t).
\end{equation}

\noindent Using the orthogonality conditions in Eq. \ref{eq:NL rom ortho}, the contribution from the time variation of the nonlinear closure vanishes under projection, resulting in:

\begin{equation}
\frac{\mathrm{d}\mathbf{q}_r}{\mathrm{d}t} =\mathbf{V}^{T}\mathbf{P}
\mathbf{f}(\tilde{\mathbf{q}},t),
\qquad
\tilde{\mathbf{q}}
=
\mathbf{q}_{\mathrm{ref}}
+
\mathbf{P}^{-1}
\left[
\mathbf{V}\mathbf{q}_r
+
\mathbf{\hat{V}}\mathcal{N}(\mathbf{q}_r)
\right],
\label{eq:NL_reduced_ode}
\end{equation}

\noindent which, similar to the quadratic nonlinear manifold, lead to a reduced dynamical system with the same dimension and the same left hand side structure as the linear Galerkin ROM.

Multiple methods have been investigated to learn the nonlinear map ($\mathcal{N}$). Previous studies \cite{barnett_neural-network-augmented_2023,cohen_nonlinear_2024} have explored the use of deep artificial neural networks (ANNs), which offer expressive representation of this complex nonlinear mapping. However, their effectiveness generally relies on sufficiently rich training dataset and appropriate selection of network architecture and hyperparameters. Moreover, the resulting mappings from ANN can be less interpretable. All of these will limit ANNs' applicability when only limited high-fidelity training data are available, which are particularly relevant to the present work with one major objective to construct effective ROMs from relatively sparse training datasets. Acoordingly, rather than employing ANN-based mappings, we consider two alternative machine-learning regression methods for constructing $\mathcal{N}$: \textbf{Gaussian Process Regression (GPR)} and \textbf{Radial Basis Function (RBF) interpolation}, both of which have been proved effective to construct interpretable nonlinear mappings while requiring comparatively limited model specification and training dataset \cite{ares_de_parga_nonlinear_2026}. In the current study, both methods are trained with identical set of training data for consistent comparison. Each training snapshot is first projected onto the primary and residual subspaces to obtain generalized coordinates for constructing the nonlinear mapping. For the $s^{th}$ training snapshot, the corresponding generalized coordinates are
\begin{equation}
\mathbf{q}_r^{\,s}
=
\mathbf{V}^{T}\mathbf{P}
\left(
\mathbf{q}^{\,s}-\mathbf{q}_{\mathrm{ref}}
\right),
\qquad
\mathbf{\hat{q}}_r^{\,s}
=
\mathbf{\hat{V}}^{T}\mathbf{P}
\left(
\mathbf{q}^{\,s}-\mathbf{q}_{\mathrm{ref}}
\right),
\label{eq:general_coord}
\end{equation}

\noindent where $s=1,\ldots,N_{\mathrm{td}}$ and $N_{\mathrm{td}}$ denotes the number of training snapshots. The resulting primary and residual generalized coordinates are collected in the training matrices
\begin{equation}
\mathbf{Q}_{r,\mathrm{td}}
=
\begin{bmatrix}
(\mathbf{q}_r^{\,1})^{T} \\
(\mathbf{q}_r^{\,2})^{T} \\
\vdots \\
(\mathbf{q}_r^{\,N_{\mathrm{td}}})^{T}
\end{bmatrix}
\in
\mathbb{R}^{N_{\mathrm{td}}\times n_p},
\qquad
\mathbf{\hat{Q}}_{r,\mathrm{td}}
=
\begin{bmatrix}
(\mathbf{\hat{q}}_r^{\,1})^{T} \\
(\mathbf{\hat{q}}_r^{\,2})^{T} \\
\vdots \\
(\mathbf{\hat{q}}_r^{\,N_{\mathrm{td}}})^{T}
\end{bmatrix}
\in
\mathbb{R}^{N_{\mathrm{td}}\times \hat{n}_p}.
\label{eq:general_coord_training_matrix}
\end{equation}

\textbf{GPR} approximates the nonlinear mapping ($\mathcal{N}$) by exploring probabilistic correlations between the primary and residual generalized coordinates in Eq.~\ref{eq:general_coord} through a prescribed kernel or covariance function. Given the training matrices in Eq.~\ref{eq:general_coord_training_matrix}, the GPR prediction of the residual generalized coordinates associated with a new primary reduced state ($\mathbf{q}_r^{*}$) is expressed as 
\begin{equation}
\mathcal{N}_{\mathrm{GPR}}
\left(
\mathbf{q}_r^{*}
\right)^{T}
=
\mathbf{K}
\left(
\mathbf{q}_r^{*},
\mathbf{Q}_{r,\mathrm{td}}
\right)
\boldsymbol{\alpha},
\label{eq:GPR_mapping}
\end{equation} 



\noindent where $\boldsymbol{\alpha}=\left[\mathbf{K}\left(\mathbf{Q}_{r,\mathrm{td}},\mathbf{Q}_{r,\mathrm{td}}\right)\right]^{-1}\mathbf{\hat{Q}}_{r,\mathrm{td}}$ represents a precomputed matrix using the training data. Here $\mathbf{K}\left(\mathbf{Q}_{r,\mathrm{td}},\mathbf{Q}_{r,\mathrm{td}}\right)$ denotes the covariance matrix of the primary generalized coordinates in the training dataset given by
\begin{equation}
\left[
\mathbf{K}
\left(
\mathbf{Q}_{r,\mathrm{td}},
\mathbf{Q}_{r,\mathrm{td}}
\right)
\right]_{ij}
=
K
\left(
\mathbf{q}_r^{\,i},
\mathbf{q}_r^{\,j}
\right),
\end{equation}
\noindent where $i,j=1,\ldots,N_{\mathrm{td}}$ denote training sample indices and $K$ is a covariance kernel describing the correlation between two states in the primary generalized coordinates. For a new primary reduced state, $\mathbf{q}_r^{*}$, its covariance with each training sample is collected in the row vector
\begin{equation}
\left[
\mathbf{K}
\left(
\mathbf{q}_r^{*},
\mathbf{Q}_{r,\mathrm{td}}
\right)
\right]_{s}
=
K
\left(
\mathbf{q}_r^{*},
\mathbf{q}_r^{\,s}
\right),
\end{equation}
\noindent where $s=1,\ldots,N_{\mathrm{td}}$. In the present work, a Mat\'ern kernel with smoothness parameter $\nu$ is employed

\begin{equation}
K
\left(
\mathbf{q}_r^{\,i},
\mathbf{q}_r^{\,j}
\right)
=
\sigma_f^2
\left(
1+
\frac{
\sqrt{3}
\left\|
\mathbf{q}_r^{\,i}-\mathbf{q}_r^{\,j}
\right\|_2
}{\ell}
\right)
\exp
\left(
-
\frac{
\sqrt{3}
\left\|
\mathbf{q}_r^{\,i}-\mathbf{q}_r^{\,j}
\right\|_2
}{\ell}
\right),
\label{eq:GPR_matern}
\end{equation}

\noindent where $\sigma_f^2$ denotes the signal variance and $\ell$ is the
characteristic length scale controlling the decay of correlation with distance
in the primary generalized coordinate. These kernel hyperparameters are determined
from the training data through marginal-likelihood optimization.

On the other hand, \textbf{RBF} interpolation provides a closely related but fully deterministic approach for constructing $\mathcal{N}$. The nonlinear map is represented as a weighted combination of radial kernels centered at the training coordinates
\begin{equation}
\mathcal{N}_{\mathrm{RBF}}
\left(
\mathbf{q}_r
\right)
=
\sum_{s=1}^{N_{\mathrm{td}}}
\mathbf{w}^{\,s}
\phi
\left(
\left\|
\mathbf{q}_r-\mathbf{q}_r^{\,s}
\right\|_2
\right),
\label{eq:RBF_mapping}
\end{equation}

\noindent where $\phi:[0,\infty)\rightarrow\mathbb{R}$ denotes the selected
radial kernel and
$\mathbf{w}^{\,s}\in\mathbb{R}^{\hat{n}_p}$ contains the interpolation weights
associated with the $s^\text{th}$ training state. The RBF kernel matrix is constructed
from the pairwise distances between the training coordinates as
\begin{equation}
\left[
\mathbf{\Phi}
\right]_{ij}
=
\phi
\left(
\left\|
\mathbf{q}_r^{\,i}-\mathbf{q}_r^{\,j}
\right\|_2
\right),
\end{equation}

\noindent where again $i,j=1,\ldots,N_{\mathrm{td}}$. The interpolation weights are
collected into the matrix
$\boldsymbol{B}\in
\mathbb{R}^{N_{\mathrm{td}}\times\hat{n}_p}$ and are computed offline by
solving,
\begin{equation}
\mathbf{\Phi}
\boldsymbol{B}
=
\mathbf{\hat{Q}}_{r,\mathrm{td}}.
\label{eq:RBF_weights}
\end{equation}

\noindent For a new reduced state $\mathbf{q}_r^{*}$, the RBF
kernel evaluations with respect to the training points are collected in
\begin{equation}
\left[
\boldsymbol{\phi}
\left(
\mathbf{q}_r^{*}
\right)
\right]_{s}
=
\phi
\left(
\left\|
\mathbf{q}_r^{*}-\mathbf{q}_r^{\,s}
\right\|_2
\right).
\end{equation}

\noindent The residual generalized coordinates are then obtained online as
\begin{equation}
\mathcal{N}_{\mathrm{RBF}}
\left(
\mathbf{q}_r^{*}
\right)^{T}
=
\boldsymbol{\phi}
\left(
\mathbf{q}_r^{*}
\right)
\boldsymbol{B}.
\label{eq:RBF_online}
\end{equation}

\noindent Common examples of RBF kernels include the Gaussian kernel 
$\phi(r)=\exp(-\epsilon^{2} r^{2})$, the multiquadric kernel 
$\phi(r)=\sqrt{1+(\epsilon r)^{2}}$, and the inverse multiquadric kernel 
$\phi(r) = (1+(\epsilon r)^{2})^{-1/2}$,
where $r$ denotes the Euclidean distance, and $\epsilon$ is a shape parameter governing the spatial locality of the basis functions. Larger values of $\epsilon$ generally produce more localized kernels, whereas smaller values lead to smoother, more globally supported basis functions. The selection of $\epsilon$ has a substantial impact on both interpolation accuracy and numerical conditioning and is typically determined through cross-validation or hyperparameter tuning. For a comprehensive discussion of machine learning regression techniques for modeling closure errors, including practical aspects of RBF interpolation and related approaches, the reader is referred to Ares De Parga et al. \cite{ares_de_parga_nonlinear_2026}.

\subsection{Hyper-reduction}
For construction of nonlinear projection-based ROMs, one major computational complexity arises from the evaluation of the high-dimensional nonlinear operator (i.e., $\mathbf{f}$ in Eq. \ref{prj.rom}), which typically scales with the system's degrees of freedom ($N$). To lift such bottleneck and achieve substantial computational acceleration, hyper-reduction techniques have been formulated to reduce the online evaluation cost of the nonlinear operator. Among these, sparse sampling methods represent the most widely adopted and well-established class, which seeks to evaluate the governing equations, or their non-linear components, at a limited set of strategically selected degrees of freedom ($n_s << N$), thereby reducing the associated computational cost. In the present work, we adopt the gappy POD formulation \cite{everson_karhunenloeve_1995} for hyper-reduction, which approximates the full-field non-linear terms through a least-squares \emph{regression} formulation by introducing oversampling into the empirical interpolation method \cite{chaturantabut_discrete_2009,chaturantabut_nonlinear_2010}. The importance of oversampling for improving stability and accuracy has been emphasized in several studies \cite{peherstorfer_stability_2020,wentland_scalable_2023}. Accordingly, in the present work, hyper-reduction is achieved using the following formulation:
\begin{equation}
\label{eq:deim}
\mathbf{\tilde{f}} \approx \mathbf{U} (\mathbf{S}^T \mathbf{U})^{\dagger} \mathbf{S}^T \mathbf{f},
\end{equation}
\noindent where $\mathbf{\tilde{f}}$ represents an approximation of the nonlinear operator $\mathbf{f}$ using selected sparse samples $\mathbf{S}^T \mathbf{f}$, and $(\mathbf{S}^T \mathbf{U})^{\dagger}$ denotes the Moore--Penrose pseudoinverse of $\mathbf{S}^T \mathbf{U}$. The matrix $\mathbf{S} \in \mathbb{R}^{N \times n_s}$ is a sampling (selection) operator composed of $n_s$ columns of the identity matrix $\mathbf{I} \in \mathbb{R}^{N \times N}$, corresponding to the selected sampling indices. For gappy POD, the number of sampling points satisfies $n_s > n_p$. The matrix $\mathbf{U} \in \mathbb{R}^{N \times n_d}$ denotes the basis employed to approximate the nonlinear term $\mathbf{f}$ and is typically constructed via POD using snapshots of $\mathbf{f}$.

\section{Adaptive projection-based model order reduction (MOR)}\label{sec:AROM}

In this section, we review the \emph{adaptive} projection-based MOR formulation employed within the proposed self-adaptive MOR framework. Specifically, we leverage the online adaptation methods introduced by Huang and Duraisamy \cite{huang_predictive_2023} and recent work by Mohaghegh and Huang \cite{mohaghegh_feature-guided_2026}, which have shown promise of incorporating adaptivity to construct truly predictive ROMs in convection-dominated fluid flow problems. The adaptive projection-based MOR seeks to update the basis ($\mathbf{V}$) and sampling points ($\mathbf{S}$) simultaneously during the online ROM calculations
\begin{equation}
\{\mathbf{V}^n,\mathbf{S}^n\} = \argmin_{\mathbf{V}^n \in \mathbb{R}^{N \times n_p} , \mathbf{S}^n \in \mathbb{R}^{N \times n_s}} \| \mathbf{V}^n(\mathbf{S}_n^T\mathbf{V}^n)^{\dagger}\mathbf{S}_n^T\mathbf{P}\mathbf{r}(\hat{\mathbf{q}}^n) \|_2^2 ,
\label{arom_main}
\end{equation}
\noindent where $\mathbf{r}({\hat{\mathbf{q}}}^n) \in \mathbb{R}^{N}$ is the fully discrete FOM equation residual defined according to the time-stepping scheme, for example, in Eq. \ref{eq:SSPRK_grouped}
\begin{equation}    \mathbf{r}(\hat{\mathbf{q}}^{n}) = \hat{\mathbf{q}}^{n} - \hat{\mathbf{q}}^{n-1} - \Delta t \sum_{i=1}^{s} \beta_i \textbf{f}(\hat{\mathbf{q}}^{(i)}, t^{(i)}),
\label{eq:fom_eq_res}
\end{equation}
\noindent with $\hat{\mathbf{q}}$ representing the estimated full states at the sampled points directly evaluated by setting the FOM equation residual to zero -- i.e., 
\begin{equation}    
\mathbf{S}_n^T \mathbf{r}(\mathbf{S}_n\mathbf{S}_n^T{\hat{\mathbf{q}}}^n + \mathbf{S}_n^*\mathbf{S}_n^{*T}{\hat{\mathbf{q}}}^n) = 0.
\label{eq:fom_eq_res_sampled}
\end{equation}
\noindent $\mathbf{S}_n^{*T} \mathbf{\hat{q}}^{n}$ represents the estimated full states at the unsampled points that are directly evaluated using gappy POD in Eq. \ref{eq:deim} - i.e., 
\begin{equation}    
\mathbf{S}_n^{*T} \mathbf{\hat{q}}^{n} =  \mathbf{S}_n^{*T} ( \mathbf{q}_{\text{ref}} + \mathbf{P}^{-1}\mathbf{V} \mathbf{\hat{q}}^n_r ),
\label{eq:fom_hyper_unsampled}
\end{equation}
\noindent where $\mathbf{\hat{q}}^n_r = (\mathbf{S}_n^T \mathbf{V}^n)^{\dagger} \mathbf{S}_n^T \mathbf{P} ( \mathbf{\hat{q}}^n - \mathbf{q}_{\text{ref}} )$. To avoid redundancy and maintain compactness of the formulation, we omit the notation for the intermediate steps $\hat{\mathbf{q}}^{(j)}$ in defining the FOM equation residual in Eq. \ref{eq:fom_eq_res}, because they follow the same form as shown in Eq. \ref{eq:SSPRK_intem}. Moreover, in the current work, we refer to Eq. \ref{arom_main} as the adaptive ROM formulation, where the basis and sampling points are both updated on the fly during the online calculations. However, directly solving Eq. \ref{arom_main} is computationally prohibitive and may be intractable and therefore we adopt a decoupled approach instead to solve two sequential minimization problems to adapt the basis and sampling points following a predictor-corrector approach:

\begin{enumerate}

\item First, the basis matrix $\mathbf{V}^{n}$ is updated from $\mathbf{V}^{n-1}$ at time step $n$ as the solution to the optimization problem,
\begin{equation}
\argmin_{\mathbf{V}^{n} \in \mathbb{R}^{N \times n_p}} 
\left\| 
\mathbf{V}^{n} (\mathbf{V}^{n-1})^{\dagger} \mathbf{\hat{Q}}^n
- 
\mathbf{\hat{Q}}^n 
\right\|_F^2,
\label{eq:basis_optimz}
\end{equation}
\noindent where $\mathbf{\hat{Q}}^n \in \mathbb{R}^{N \times w}$ denotes a data matrix collecting centered and normalized estimated full states $\mathbf{\hat{q}}'^{n}$ over a temporal window of size $w$,
\begin{equation}
\mathbf{\hat{Q}}^n 
= 
\left[
\mathbf{\hat{q}}'^{\,n-w+1}, 
\dots, 
\mathbf{\hat{q}}'^{\,n-1}, 
\mathbf{\hat{q}}'^{\,n}
\right]
\in \mathbb{R}^{N \times w}.
\label{data_window}
\end{equation}
Where $\mathbf{\hat{q}}'^{n} = \mathbf{P} ( \mathbf{\hat{q}}^n - \mathbf{q}_{\text{ref}} )$. A feasible closed-form solution consistent with the minimization problem in Eq.~\ref{eq:basis_optimz} is given by
\begin{equation}
\mathbf{V}^{n} 
= 
\mathbf{\hat{Q}}^n 
\left[
(\mathbf{V}^{n-1})^{\dagger}
\mathbf{\hat{Q}}^n
\right]^{\dagger}.
\label{dire_adapt}
\end{equation}

\item Second, once the basis has been updated according to Eq.~\ref{eq:basis_optimz}, the sampling operator is adapted from $\mathbf{S}_{n-1}$ to $\mathbf{S}_n$. Following similar strategies as in Refs.~\cite{uy_reduced_2022,peherstorfer_model_2020}, the objective is to minimize the interpolation error associated with the gappy POD reconstruction of the estimated full state $\mathbf{\hat{q}}^{n}$ from sparse samples,
\begin{equation}
\mathbf{S}_n 
= 
\underset{\mathbf{S}_n \in \mathbf{S}^{N_s}}{\arg\min} 
\left\|
\mathbf{\hat{q}}^{n} 
- 
\mathbf{\bar{\hat{q}}}^{\,n}
\right\|_2^2,
\label{eq:interp_error}
\end{equation}
\noindent where
\[
\mathbf{\bar{\hat{q}}}^{\,n} 
= \mathbf{q}_{\text{ref}} + \mathbf{P}^{-1} \mathbf{V}^n \left( \mathbf{S}_n^T \mathbf{V}^n \right)^{\dagger} \mathbf{S}_n^T \mathbf{P} ( \mathbf{\hat{q}}^n - \mathbf{q}_{\text{ref}} ),
\]
and $\mathbf{V}^n$ denotes the updated basis from Eq.~\ref{eq:basis_optimz}. In the current work, we pursue a recently developed \emph{feature-guided sampling} (FGS) method \cite{mohaghegh_feature-guided_2026} that has shown particular promise in constructing predictive adaptive ROM for shock-dominated problems. The FGS method seeks to strategically place sampling points at locations corresponding to regions with high gradients of representative features, $|\nabla \boldsymbol{\uptheta}|$, computed from estimated future full states ($\mathbf{\hat{q}}^{\,n+z_s}$). Specifically, the sampling operator is determined as
\begin{equation}
\mathbf{S}_n 
= 
\underset{\mathbf{S}_n \in \mathbf{S}^{n_s}}{\arg\max} 
\left\|
\nabla \boldsymbol{\uptheta}^n
\right\|_2^2,
\label{eq:fgs_def}
\end{equation}
\noindent where 
\[
\boldsymbol{\uptheta}^n 
= 
\mathbf{\Theta}(\mathbf{\hat{q}}^{\,n+z_s}),
\]
and $\mathbf{\Theta} : \mathbb{R}^N \rightarrow \mathbb{R}^N$ denotes a nonlinear operator that extracts target features from the predicted future full states ($\mathbf{\hat{q}}^{\,n+z_s}$). As demonstrated in our previous work \cite{mohaghegh_feature-guided_2026}, the FGS method is conceptually related to eigenvector-based sampling method \cite{peherstorfer_stability_2020} , but adopts a direct physics-informed strategy by explicitly targeting dynamically significant features inferred from predicted states. The effectiveness of FGS method therefore depends on the appropriate selection of the feature variable $\boldsymbol{\uptheta}$, which requires \emph{a priori} physical knowledge of the governing dynamics in the system. For example, pressure ($p$) provides a natural indicator for identifying shock-dominated dynamics, whereas temperature ($T$) is generally more appropriate for characterizing flame-dominated dynamics. Finally, we remark that the adaptation of the sampling points requires evaluating the full states ($\mathbf{\hat{q}}^{\,n+z_s}$) at all the points, which
incurs high computational costs that scale with the total number of degrees of freedom ($N$) in the FOM. Therefore, the sampling points are adapted every $z_s$ time steps following the strategy in our previous works~\cite{mohaghegh_feature-guided_2026,huang_predictive_2023}. It is evident that $z_s$ is a crucial
parameter in both the adaptation of the basis and the sampling points, affecting the accuracy and  limiting the efficiency of the adaptive ROM~\cite{mohaghegh_feature-guided_2026,huang_predictive_2023}.

\end{enumerate}

\section{Motivating example for self-adaptive MOR}\label{sec:motiv_SAROM}

\begin{figure}[pos=ht] 
    \centering 
    \includegraphics[width=0.9\textwidth]{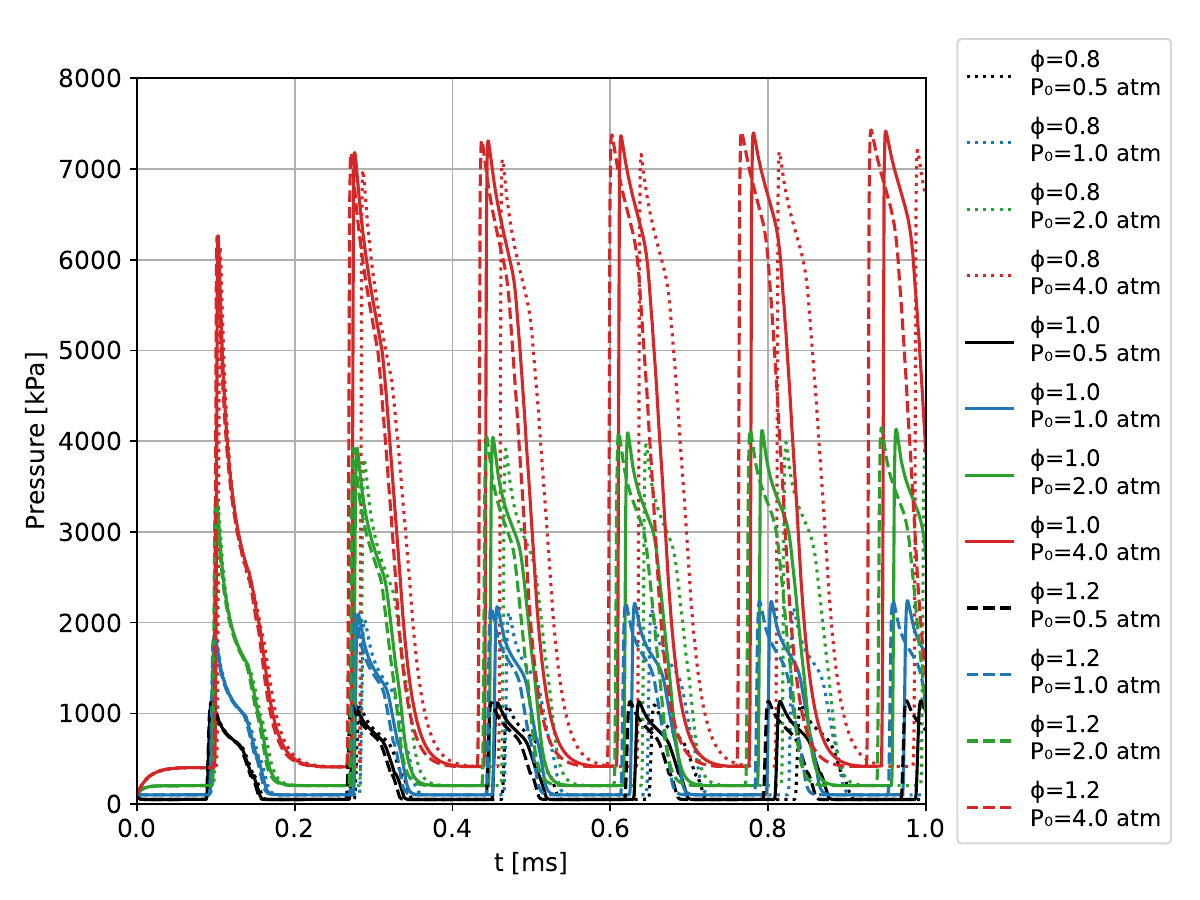} 
    \caption{Comparisons of pressure time traces at 12 different operating conditions.} 
    \label{fig:P_trace_fom_param} 
\end{figure}

Before introducing the formulations of the proposed self-adaptive projection-based MOR framework, we first illustrate the key challenge that motivates this new development using a baseline unsteady FOM simulation of a one-dimensional (1D) rotating detonation engine (RDE) with the details in model equations and configurations presented later in Sec. \ref{sec:1D_RDE}. Building upon this baseline configuration, a parametric dataset is generated by performing FOM simulations at 12 different operating conditions. Each simulation is advanced in time from \(t = 0\) to \(t = 2.5\ \text{ms}\), allowing a detonation wave to develop and establish sufficiently long stationary statistic behaviors within the computational domain - i.e.., repetitive detonation wave traveling through the domain. Figure \ref{fig:P_trace_fom_param} compares the pressure time traces monitored at the left corner of the domain for all 12 operating conditions \(t = 0\ \text{ms}\) and \(t = 1\ \text{ms}\), which exhibit distinctly different detonation-wave characteristics, including wave intensity and speed, under variable operating conditions. 

\begin{figure}[pos=ht] 
    \centering 
    \includegraphics[width=0.7\textwidth]{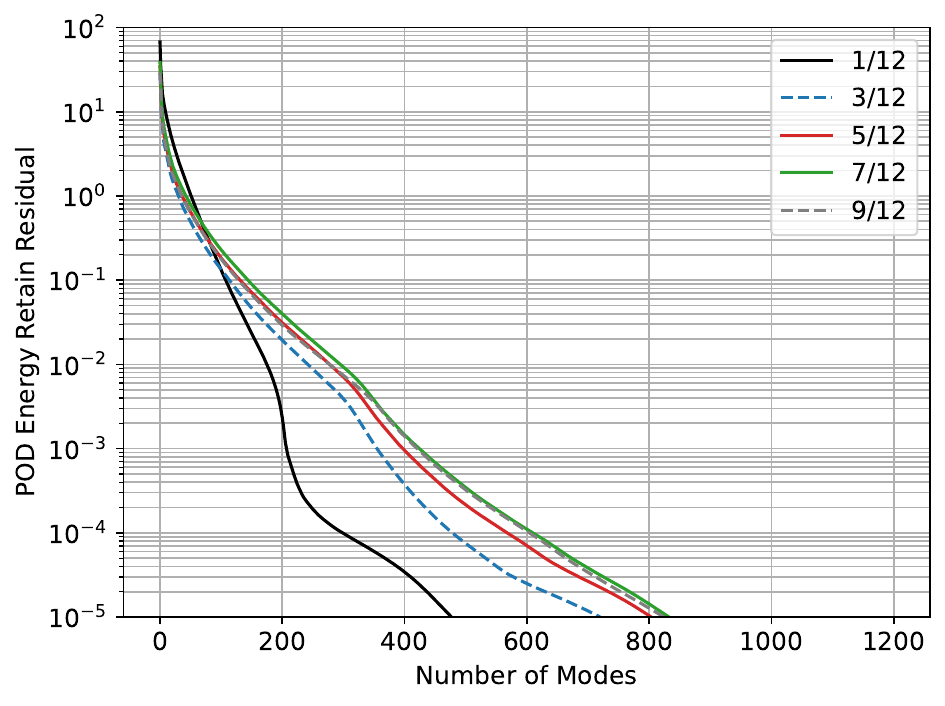} 
    \caption{POD residual energy distribution with different sizes of training dataset included for construction of POD trial basis} 
    \label{fig:pod_res} 
\end{figure}

Next, we examine the characteristics of conventional static ROM with linear subspace (as described in Sec. \ref{sec:SROM}) for this RDE test problem by progressively enriching the training dataset used to construct the POD trial basis. The analysis starts with a trial basis constructed by including the data from a single condition among the total 12 available ones, and incrementally enriches the training dataset by adding data from two additional conditions at each subsequent step, which lead to 5 sets of POD bases. The characteristics of the resulting bases are first assessed to examine how effectively these bases can capture the detonation-wave dynamics contained in the complete FOM dataset (i.e., all 12 conditions), which is quantified using the POD residual energy defined as

\begin{equation}
\text{POD Residual Energy}(n_p), \% = \left(1 - \frac{\sum_{i=1}^{n_p} \tilde{\sigma}_i^2}{\sum_{i=1}^{n_{p,\text{total}}} \tilde{\sigma}_i^2}\right) \times 100 \%
\label{eq:POD_res}
\end{equation}

\noindent where $\tilde{\sigma}_i$ denotes the $i^{th}$ singular value from the SVD used to construct the trial basis ($\mathbf{V}$) in descending order. Here, $n_p$ is the number of modes retained in the POD trial basis, and $n_{p,\text{total}}$ is the total number of modes available in the training dataset. The resulting POD residual-energy distributions are presented in Fig.~\ref{fig:pod_res}. It is readily observed that a relatively large number of POD modes is required to achieve a sufficiently low residual energy. For example, at least 187 modes are needed to recover 99.99\% of the total snapshot energy. Moreover, as the training dataset is progressively enriched, the number of POD modes required to achieve the same energy threshold increases accordingly, reflecting the increasing dimensionality of the dynamics represented by the expanded dataset. These observations are characteristic of the slow decay of the Kolmogorov N-width commonly encountered in convection-dominated problems. On the other hand, the residual-energy distributions exhibit insignificant additional change once more than 5 of the 12 operating conditions are included in the training dataset. This convergence suggests that the corresponding POD trial basis have become sufficiently representative of the dominant dynamics contained within the target FOM dataset.

\begin{figure}[pos=ht] 
    \centering 
    \includegraphics[width=\textwidth]{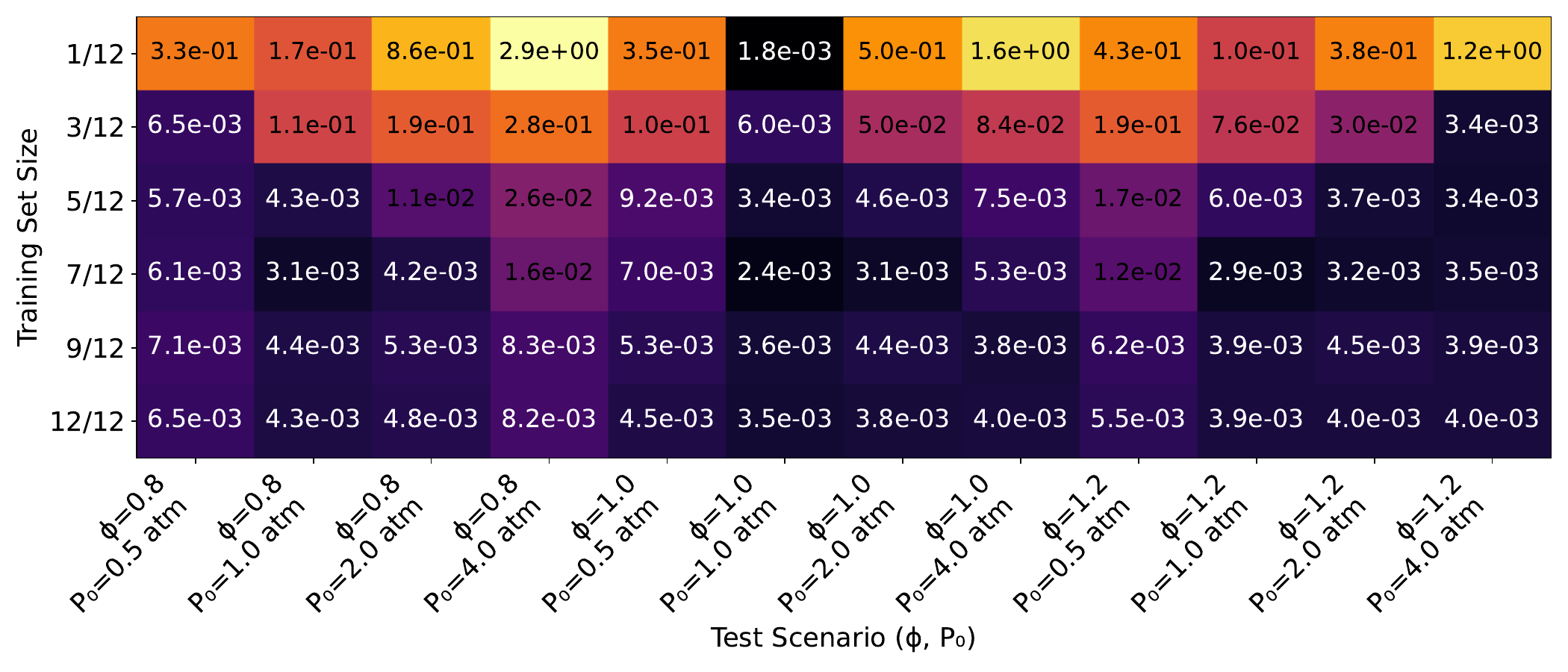} 
    \caption{Heat map of static ROMs' prediction errors with different sizes of training dataset included for ROM construction} %
    \label{fig:heatmap} 
\end{figure}

Furthermore, individual static ROMs are constructed using the generated POD bases corresponding to different sizes of training dataset in Fig. \ref{fig:pod_res}. Specifically, the number of POD modes recovering 99.9\% of total energy are used to construct each static ROM. The predictive capabilities of the resulting ROMs are then evaluated across all 12 operating conditions using the time- and variable-averaged relative $L_2$-norm errors Fig.\ref{fig:heatmap},
\begin{equation}
   \overline{\epsilon}_{total} 
   = \frac{1}{k} \sum_{n=1}^{k} 
     \left[
        \frac{1}{N_{\text{var}}}
        \sum_{i=1}^{N_{\text{var}}}
        \frac{\|\mathbf{q}_{FOM,i}^n - \mathbf{q}_{ROM,i}^n\|_2^2}
             {\|\mathbf{q}_{FOM,i}^n\|_2^2}
     \right],
   \label{eq:total_err_time_avg}
\end{equation}

\noindent where ${\mathbf{q}}_{FOM,i}^n$ and ${\mathbf{q}}_{ROM,i}^n$ represent the $i^{th}$ state variable solution of FOM and ROM at time step $n$, respectively. Figure \ref{fig:heatmap} summarizes the prediction errors of the resulting static ROMs across all 12 operating conditions are summarized in the form of a heat map. Consistent with the analysis on POD residual energy in Fig. \ref{fig:pod_res}, the predictive accuracy of the static ROMs improves systematically as additional operating conditions are incorporated into the training dataset. Reasonably accurate predictions, characterized here by errors of approximately $1\times10^{-3}$, are achieved only after more than 5 of the 12 available operating conditions are included in the training dataset. Further enrichment beyond this point produces progressively smaller improvements, consistent with the convergence observed in the POD residual-energy distributions. Importantly, these results demonstrate that constructing a reliable and sufficiently representative static ROM requires high-fidelity training data covering a substantial fraction of the target operating conditions. In the present example, more than 5 of the 12 available FOM conditions must be included before consistently accurate predictions are obtained. For large-scale applications in which each FOM simulation is computationally expensive, such extensive coverage of the parameter space can impose a considerable offline data-generation cost, highlighting a fundamental limitation of conventional MOR approaches based on a global static trial basis.

These observations strongly motivate the development of a more robust ROM framework that can substantially reduce or eliminate the extensive offline training costs associated with conventional static ROM approaches while maintaining computational efficiency during online prediction. As discussed in the preceding sections, adaptive MOR method provides a promising pathway toward this objective by acquiring information and updating the reduced representation as previously unseen dynamics are encountered. Such online adaptation strategy can substantially reduce the dependence of ROM performance on offline training dataset while inherently improving predictive robustness beyond the training regime. However, this capability comes at an additional computational cost due to the operations associated with frequent basis and sampling updates, which can make adaptive ROMs considerably more expensive than conventional static ROMs. To balance this trade-off between accuracy and efficiency, the present work introduces a \emph{self-adaptive projection-based MOR framework} that dynamically transitions between adaptive and static ROM formulations based on the characteristics of the evolving dynamics. 

\section{Self-adaptive projection-based MOR framework}\label{sec:SAROM}

In this section, we introduce the detailed formulation of the proposed self-adaptive projection-based MOR framework, which incorporates both the predictive capability of adaptive ROMs for previously unseen dynamics and the computational efficiency of conventional static ROMs. The primary objective is to establish a fully automated and systematic mechanism that transition between adaptive and static ROM formulations according to the evolution of the underlying system dynamics. The central idea is to deploy adaptive ROM only when the existing reduced representation is insufficient to capture newly encountered dynamics and to transition to a computationally more efficient static ROM once the acquired information provides an adequate representation of the subsequent dynamics. Specifically, the proposed self-adaptive framework consists of three primary stages and the complete computational procedure of implementing the self adaptive MOR framework is summarized in Algorithm \ref{alg:main}:

\begin{enumerate}
    \item {\emph{Execute adaptive ROM with in-situ basis monitoring}} -- The self-adaptive framework is initialized in the adaptive-ROM mode with two POD bases $\mathbf{V}^n$ and $\mathbf{V}^n_\mathrm{inc}$. Here $\mathbf{V}^n$ represents the primary adaptive-ROM basis, which evolves at each time step according to Eq.\ref{arom_main}, whereas $\mathbf{V}^n_\mathrm{inc}$ represents a memory basis that accumulates information (or snapshots) from previously encountered dynamics. Specifically, newly available snapshots are incorporated into $\mathbf{V}^n_\mathrm{inc}$ at a prescribed time step $z_i$ using incremental singular value decomposition (incSVD) \cite{goos_incremental_2002,zhang_answer_2022,kuhl_incremental_2024}, while retaining only the leading the $n_p$ POD modes. Consequently, $\mathbf{V}_\mathrm{inc}^n$ provides a compact representation of the dynamics accumulated throughout the simulation, whereas $\mathbf{V}^n$ represents the dynamics locally encountered by the adaptive ROM. Next, to determine whether the currently encountered dynamics have already been sufficiently represented by the accumulation information, the framework evaluates the \emph{principal angles} between $\mathbf{V}^n$ and $\mathbf{V}^n_\mathrm{inc}$, which provide a geometric measure of similarity between two vector subspaces and can be efficiently computed from the singular values of the inner product between their orthonormal basis matrices \cite{knyazev_principal_2002}. Specifically, given 
    $\mathbf{V}^n_\mathrm{inc}$ and $\mathbf{V}^{n}$ $\in \mathbb{R}^{N \times n_p}$ with orthonormal columns, the singular values obtained from the reduced SVD of $(\mathbf{V}^n_\mathrm{inc})^T \mathbf{V}^{n}$ satisfy:
    \begin{equation}
    1 \ge \sigma_{1} \ge \sigma_{2} \ge \cdots \ge \sigma_{n_p} \ge 0.
    \label{eq:svd_principal_angles}
    \end{equation}
    
    \noindent The corresponding principal angles are defined as
    \begin{equation}
    \phi_{k} = \arccos(\sigma_{k}), 
    \qquad k = 1, \ldots, n_p,
    \end{equation}
    such that
    \begin{equation}
    0 \le \phi_{1} \le \cdots \le \phi_{n_p} \le \frac{\pi}{2}.
    \end{equation}
    
    \noindent A scalar measure of the overall subspace discrepancy at time step $n$ is then defined using the root mean square (RMS) of the principal angles as
    \begin{equation}
    \label{eq:subspace_rms}
    \Phi_{n} = \sqrt{\frac{\sum_{k=1}^{n_p} \phi_k^2}{n_p}} .
    \end{equation}
    
    In addition to the instantaneous subspace discrepancy $\Phi_{n}$, its temporal variation over the monitoring interval is also examined by
    \begin{equation}
    \label{eq:subspace_tempVar}
    \Phi_{n}^{'} = |\Phi_{n} - \Phi_{n-z_i}|.
    \end{equation}
    
    Together, $\Phi_{n}$ and $\Phi_{n}^{'}$ quantify whether the current adapting subspace is sufficiently represented by the accumulated memory basis and whether this relationship has stabilized over time. When
    \begin{equation}
    \Phi_n < \varepsilon_\Phi,
    \quad \text{and} \quad
    \Phi^{'}_n < \varepsilon_{\Phi^{'}},
    \label{eq:subspace_criteria}
    \end{equation}
    are simultaneously satisfied for prescribed tolerances 
    $\varepsilon_\phi,\, \varepsilon_{\Phi^{'}} > 0$, the framework identifies that the adaptive ROM is predominantly encountering dynamics that have already been represented by the accumulated solution history. This condition triggers consideration of a transition from the adaptive-ROM mode to the static-ROM mode. 

\item{\emph{Construct and assess static ROM on the fly}} -- Once the conditions in Eq. \ref{eq:subspace_criteria} are satisfied, the framework proceeds to construct an effective static ROM to model the future-state dynamics. To achieve this, a separate POD basis $\mathbf{V}_{\mathrm{inc}}^{'n}$ and its corresponding singular values $\boldsymbol{\Sigma}_{\mathrm{inc}}^{'n}$ are computed and continuously enriched using new information via incremental SVD. Unlike $\mathbf{V}_{\mathrm{inc}}^n$, whose dimension is truncated to $n_p$ for monitoring purposes, the candidate static basis is incrementally enriched as additional snapshots become available, allowing its dimension to increase until the encountered dynamics are represented to a prescribed accuracy. Specifically, every newly available high-fidelity snapshot $\mathbf{\tilde{q}}^{n}$ is incorporated at a prescribed update interval $z_j$ to update $\mathbf{V}_{\mathrm{inc}}^{'n}$ and $\boldsymbol{\Sigma}_{\mathrm{inc}}^{'n}$ via
    \begin{equation}
    \label{eq:incSVD}
    \{\mathbf{V}_{\mathrm{inc}}^{'n},\, \boldsymbol{\Sigma}_{\mathrm{inc}}^{'n}\}
    =
    \operatorname{incSVD}\!\left(
    \mathbf{V}^{'n-z_j}_{\mathrm{inc}},
    \boldsymbol{\Sigma}^{n-z_j}_{\mathrm{inc}},
    \mathbf{\tilde{q}}^{n}
    \right).
    \end{equation}
    
    \noindent Furthermore, the representativeness of the continuously updated candidate basis is simultaneously quantified by computing the projection error ($\overline{\epsilon}_{proj}^n$) of each newly arriving snapshot, which is defined based on the variable-averaged $L_2$-norm error, 
    \begin{equation}
   \overline{\epsilon}_{proj}^n 
   =    \frac{1}{N_{\text{var}}}
        \sum_{i=1}^{N_{\text{var}}}
        \frac{\|(\mathbf{I}-\mathbf{V}_{\mathrm{inc}}^{'n}{\mathbf{V}_{\mathrm{inc}}^{'n}}^T) \mathbf{\tilde{q}}^n\|_2^2}
             {\|\mathbf{\tilde{q}}^n\|_2^2}.
       \label{eq:proj_error_isvd}
    \end{equation}

    \noindent  When the projection error falls below a prescribed thresholds $\epsilon_{proj}$ -- i.e.,
    \begin{equation}
    \overline{\epsilon}_{proj}^n < \epsilon_{proj},
    \label{eq:proj_error_criterion}
    \end{equation}
    the newly arrived snapshots contribute negligible information outside the candidate subspace, indicating that the constructed basis has reached sufficient enrichment for the dynamics encountered during the adaptive-ROM mode. We also remark that incremental SVD procedure itself requires evaluation of the component of each incoming snapshot orthogonal to the existing basis \cite{zhang_answer_2022} and therefore, the information required to evaluate this projection-error criterion in Eq. \ref{eq:proj_error_criterion} is already available as part of the basis-update procedure and can be reused with negligible additional computational overhead. Then the resulting trial basis is used to construct the static ROM following the formulations in Sec. \ref{sec:SROM}.
    
    \item \emph{Execute the static ROM} -- Once the subspace-monitoring criteria in Stage 1 indicate that the evolving dynamics have become sufficiently recurrent and the projection-error criterion in Stage 2 confirms that the candidate static basis is sufficiently enriched, the framework transitions from the adaptive-ROM mode to the static-ROM mode. The adaptive basis updates and associated sampling operations are then deactivated, and the static ROM constructed during Stage 2 is employed for subsequent time integration. In this manner, the framework retains the predictive capability of online adaptation while it is needed to acquire previously unseen dynamics, but eliminates the associated adaptation overhead once those dynamics can be adequately represented by the constructed trial basis.

\end{enumerate}

The proposed self-adaptive MOR framework offers two key advantages. \emph{First}, it substantially reduces the need for extensive \emph{a priori} FOM training data by allowing the low-dimensional representation to be constructed and refined online as new dynamics are encountered. \emph{Second}, instead of constructing a single global static basis that must simultaneously represent dynamics across the entire parameter space, the framework generates scenario-specific subspaces tailored to the dynamically relevant features encountered during each simulation. Such localized representations can require substantially fewer basis vectors than a global basis assembled from snapshots spanning many operating conditions, including dynamics that may be irrelevant to the particular scenario being predicted. Once an effective scenario-specific representation has been acquired, transitioning to the static ROM allows this information to be reused without continuously incurring the computational overhead associated with online adaptation.

\begin{algorithm}
\caption{Self Adaptive ROM Framework}
\label{alg:main}
\begin{algorithmic}[1]

\State \textbf{Input}: $w_{init}$, $\epsilon_{\Phi_{n}}$, $\epsilon_{\Phi_{n}^{'}}$, $\epsilon_{proj}$,$z_i$, ,$z_j$, $M$

\State Adaptive ROM training mode (FOM)     = True
\State Adaptive ROM with basis monitoring   = False
\State Adaptive ROM with basis construction = False
\State Execute Static ROM                   = False

\For{n=$0,...,M$}

\If{Adaptive ROM training mode (FOM)}
    \State Advance solution using FOM (Eq. \ref{eq:fom})
    \If{$n=w_{init}$}
        \State Initialize adaptive ROM parameters: $\mathbf{V}^n$, $\mathbf{S}^n$, $\mathbf{q}_{\text{ref}}$, $\mathbf{P}$.
        \State Initialize monitoring metrics: incremental SVD $\mathbf{V}^n_\mathrm{inc}$, subspace angle $\Phi_{n}$ and it's gradient $\Phi_{n}^{'}$.
        \State Adaptive ROM training mode (FOM) = False
        \State Adaptive ROM with basis monitoring = True
    \EndIf

\ElsIf{Adaptive ROM with basis monitoring}
    \State Advance solution using Adaptive ROM (Eq. \ref{arom_main})
    \If{$mod(n,z_{i})==0$}
    \State Append new solution $\mathbf{\tilde{q}^n}$ to compute $\mathbf{V}^n_\mathrm{inc}$ 
    \State Calculate the RMS of principal angles $\mathbf{\Phi}^n$ (Eq. \ref{eq:subspace_rms}) between $\mathbf{V}^n_\mathrm{inc}$ and $\mathbf{V}^n$
    \State Compute the temporal variation of principal angle $\Phi_{n}^{'} = |\Phi_{n} - \Phi_{n-z_i}|$
    \If{    $\Phi_n < \varepsilon_\Phi$, $\Phi^{'}_n < \varepsilon_{\Phi^{'}}$ }
        \State Adaptive ROM with basis monitoring   = False
        \State Adaptive ROM with basis construction = True
        \State Initialize static basis incremental SVD: $\mathbf{V}_{\mathrm{inc}}^{'n}$ and $\boldsymbol{\Sigma}_{\mathrm{inc}}^{'n}$
    \EndIf
    \EndIf
\ElsIf{Adaptive ROM with basis construction}
    \State Advance solution using Adaptive ROM (Eq. \ref{arom_main})
    \If{$mod(n,z_j)==0$}
        \State Update basis $\mathbf{V}_{\mathrm{inc}}^{'n}$ via incremental SVD (Eq. \ref{eq:incSVD}) and store the the projection error. 
        \If{$\overline{\epsilon}_{proj}^n < \epsilon_{proj}$}
            \State Adaptive ROM with basis construction = False
            \State Execute Static ROM = True
            \State Perform resampling with new static basis $\mathbf{V}_{\mathrm{inc}}^{'n}$
        \EndIf
    \EndIf

\ElsIf{Execute Static ROM}
    \State Advance solution in time using the selected static ROM strategies described in Sec. \ref{sec:SROM}
\EndIf

\EndFor

\end{algorithmic}
\end{algorithm}

\section{Numerical results and analysis}
\label{sec:results}

To assess the performance and capabilities of the proposed self adaptive MOR framework on complex multi-physics problems, especially those featuring shock-dominated dynamics, we consider a suite of rotating detonation engine (RDE) benchmark test problems, which are characterized by strongly transient and convection-dominated dynamics. These test problems involve the formation and propagation of detonation waves governed by tightly coupled shock waves and chemical reactions, leading to sharp gradients and discontinuities that pose significant challenges for conventional projection-based ROMs. The first benchmark is a one-dimensional (1D) premixed RDE problem developed by the current authors to reproduce representative multi-physics detonation-wave dynamics within a computationally tractable 1D configuration. This benchmark problem enables a detailed assessment of the predictive accuracy, robustness, and computational efficiency of the proposed framework for challenging shock-dominated problems while avoiding the prohibitive cost usually required for these high-dimensional simulations. More importantly, it provides an ideal testbed for prototyping, developing, and improving data-driven ROM methodologies for complex multi-physics problems, which benefits a broader community of computational science. The second benchmark considers a two-dimensional (2D) premixed RDE configuration with similar dominant detonation-wave characteristics of the 1D problem while introducing additional multi-dimensional flow features and turbulence. This increased physical complexity and computational scale make this problem substantially more challenging for ROM development. Accordingly, the 2D benchmark is used to further evaluate the performance of the proposed framework, with particular emphasis on its scalability to large-scale and multi-dimensional systems.

\subsection{One-dimensional (1D) premixed RDE}
\label{sec:1D_RDE}
First, we consider a 1D premixed RDE problem, which simulates the detonation-wave dynamics in the azimuthal direction and has a computational domain with a length of 287 mm, discretized with 2000  uniform finite volume cells -- i.e., $\Delta x = 0.1435 \ \text{mm}$. Periodic boundary conditions are specified at the two ends of the domain to simulate the traveling of the detonation wave around the azimuthal direction. To account for the influence of the inlet injection, a simplified quasi-injection model is utilized with variable inflow conditions applied, based on injector conditions $(p, T, v)$, assuming isentropic expansion through the injector nozzle into the combustion chamber. Details of this injection model is shown in Appendix B. 

The combustion process is modeled using a detailed $H_2$--$O_2$ chemical kinetic mechanism \cite{oran_weak_1982}, which consists of 10 species and 24 elementary reactions. An in-house 1D CFD solver\footnote{\url{https://github.com/alimike97/CompFlowLab}} has been developed and employed for both the FOM and ROM calculations of this 1D RDE problem. The solver supports chemically reactive flow simulations through integration of the open-source thermochemistry library Cantera \cite{speth_canteracantera_2017} and solves the fully coupled conservation equations for mass, momentum, energy, and species transport. Further details on the governing equations are provided in Appendix A. Spatial discretization is performed using a cell-centered, second-order accurate finite volume method. Inviscid fluxes are evaluated using the Roe scheme \cite{roe_approximate_1981}, and flow field monotonicity in the presence of strong gradients is enforced through the Barth--Jespersen gradient limiter \cite{barth_design_1989}, with a ghost-cell treatment applied at the boundaries. Time integration is carried out using an explicit four-stage, third-order strong-stability-preserving Runge--Kutta scheme (SSP-RK3) \cite{durran_numerical_2010}. All FOM and ROM computations reported in this section are performed on a single compute node equipped with dual Intel Xeon 6240R CPUs and $192\,\text{GB}$ of DDR4-2933 memory.

\begin{figure}[pos=ht] 
    \centering 
    \includegraphics[width=0.7\textwidth]{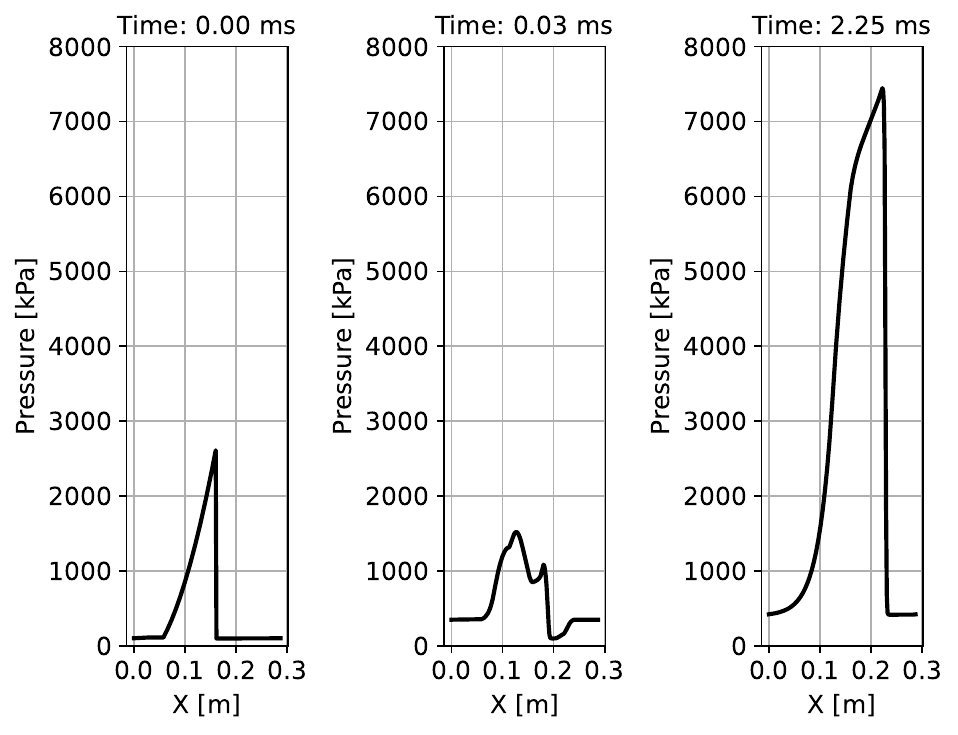} 
    \caption{Representative instantaneous FOM solutions from 1D RDE at the baseline condition of \(\phi = 1.0\) and  \(P_{in} = 4 \, atm\)} 
    \label{fig:fom_rep} 
\end{figure}


\subsubsection{Representative FOM results}
The baseline unsteady FOM simulation of the RDE configuration is performed with an operating condition of \(\phi = 1.0\) and  \(P_{in} = 4 \, atm\). To mitigate numerical instabilities during the simulation startup, a ramp-type initial condition is prescribed as shown in Fig. \ref{fig:fom_rep} at \(t = 0\ \text{ms}\). Specifically, the pressure and temperature fields increase linearly along the domain, followed by a sharp discontinuity introduced at an intermediate location, which initiates the formation of a detonation wave that propagating from the left to the right. The solution is then advanced from \(t = 0\) to \(t = 2.5\ \text{ms}\), which undergoes a transient startup phase to form a detonation wave (Fig. \ref{fig:fom_rep} at \(t = 0.03\ \text{ms}\)) and eventually establishes long-term statistically stationary and repetitive detonation-wave dynamics (Fig. \ref{fig:fom_rep} at \(t = 2.25\ \text{ms}\)) within the computational domain. 

Building on this baseline configuration, a parametric FOM dataset is generated for a total of 12 different operating conditions, which cover three equivalence ratios, \(\phi = 0.8,\ 1.0, 1.2\) and four inlet pressures, \(P_{\text{in}} = 0.5,\ 1,\ 2,\ 4 \) atm. All FOM simulations are intialized with the same prescribed state in Fig. \ref{fig:fom_rep} at \(t = 0\ \text{ms}\). The resulting parameter space produces substantial variations in the detonation-wave dynamics, including differences in wave intensity, propagation speed, and overall flow-field structure as summarized in the comparisons of all the 12 local pressure time traces between \(t = 0\ \text{ms}\) and \(t = 1\ \text{ms}\) in Fig. \ref{fig:P_trace_fom_param}. It can be seen that overall the wave intensity (indicated by the pressure magnitude) gets amplified with the increase in inlet pressure and equivalence ratio, which reflects the detonation strengthening under higher injecting pressure and richer reactant mixtures. In addition, despite all simulations starting from the same initial condition, noticeable differences in the instantaneous detonation-front locations (indicated by the abrupt pressure-magnitude increase in Fig. \ref{fig:P_trace_fom_param}) also emerge at different operating conditions. This spatial shifts arise from variations in the detonation-wave speed across the parameter space: cases associated with higher injection pressures and richer reactant mixtures produce hotter detonation and therefore higher wave speeds and vice versa. As a result, cases with faster detonation-wave speeds travel farther through the periodic domain over a given time interval, leading to distinct instantaneous front locations. Consequently, the parametric dataset encompasses not only substantial variations in the magnitude of the flow variables but also significant spatial variations in the dominant propagating features, providing a challenging test environment for evaluating the predictive capabilities of the proposed self-adaptive MOR framework.

\subsubsection{Assessment of self-adaptive ROM with different static-ROM construction strategies}

First, we evaluate the performance of self-adaptive ROM using the baseline configuration with $\phi = 1$ and $P_{in} = 4\,\text{atm}$, corresponding to the stoichiometric condition that leads to the strongest detonation-wave dynamics and therefore presents the most challenges for ROM development. Specifically, we focus on examining the self-adaptive ROM's performance using three different strategies to construct the static ROM as introduced in Sec. \ref{sec:SROM}. Each self-adaptive ROM simulation begins in the adaptive-ROM mode, initialized with 10 time step FOM calculation (i.e., $w_\text{init} = 10$), and constructed using 10 POD modes (i.e., $n_p = 10$) and 20 sample points (i.e., $n_s = 20$, corresponding to $1\%$ of the total number of cells), with the basis and sample points updated every 10 time steps ($z_s = 10$). The first 10 of the total 20 sampling points are selected using QDEIM \cite{drmac_new_2015} while the remaining 10 are determined using the recently developed FGS strategy \cite{mohaghegh_feature-guided_2026}, within which the pressure gradients are employed as the feature indicator for tracking the dynamically evolving shocks and detonation waves.

\begin{figure}[pos=ht] 
    \centering 
    \includegraphics[width=0.7\textwidth]{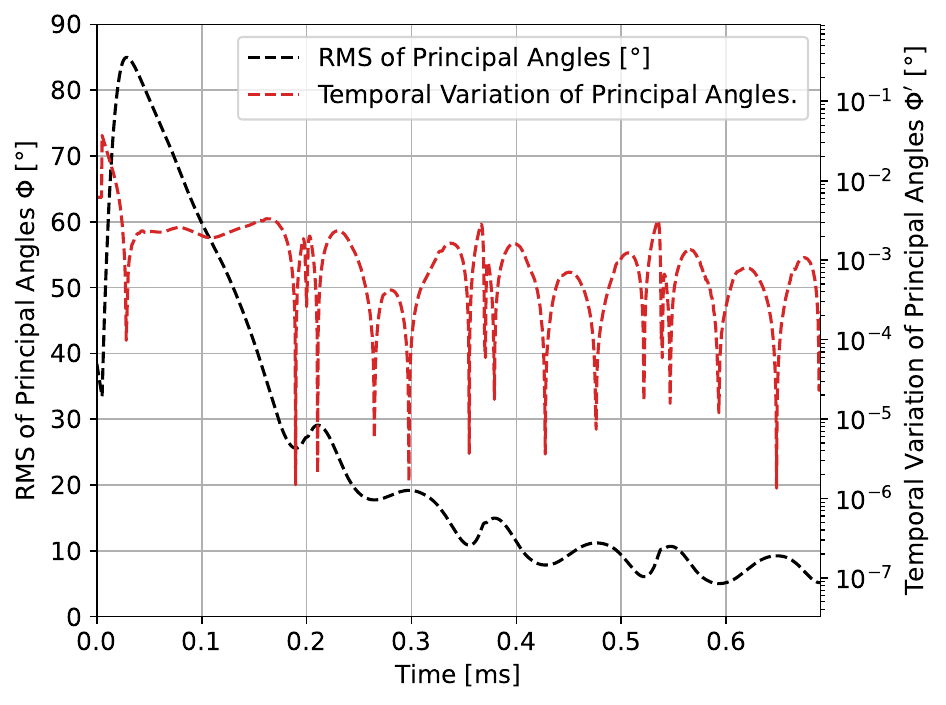} 
    \caption{Evolution of principal angles during basis monitoring stage.} 
    \label{fig:subspace_angles} 
\end{figure}

Following the three-stage computational procedure introduced in Sec. \ref{sec:SAROM}, at \textbf{Stage 1}, each self-adaptive ROM simulation begins with the adaptive-ROM mode while continuously monitoring the evolution of principal angles between the primary adapting ($\mathbf{V}^n$) and memory ($\mathbf{V}^n_\mathrm{inc}$) bases, which is measured using both their RMS ($\Phi_{n}$) and temporal variation ($\Phi^{'}_{n}$) defined in Eqs. \ref{eq:subspace_rms} and  \ref{eq:subspace_tempVar}, respectively. The corresponding tolerances are prescribed as $\varepsilon_\phi = 10^\circ$ and $\varepsilon_{\Phi^{'}} = 1\times10^{-6}$. As shown in Fig.~\ref{fig:subspace_angles}, both principal-angle metrics exhibit pronounced initial spikes associated with the rapidly evolving dynamics during the RDE startup phase, followed by a gradual stabilization as the detonation-wave dynamics become increasingly recurrent. At $t = 0.65\,\text{ms}$, both $\Phi$ and $\Phi'$ satisfy their prescribed tolerances, indicating that the $\mathbf{V}^n$ is sufficiently similar to $\mathbf{V}^n_\mathrm{inc}$ (i.e., the encountering dynamics can be well represented by the accumulated solution history) and thereby satisfying the Stage 1 criteria for consideration of adaptive-to-static transition. During \textbf{Stage 2}, the simulation continues to advance using the adaptive ROM \emph{while} a separate candidate POD basis $\mathbf{V}_\mathrm{inc}^{' n}$ is concurrently constructed via incremental SVD with newly acquired solution snapshots incorporated into the basis construction and the corresponding projection error ($\overline{\epsilon}_{proj}^n$) computed and monitored. At $t = 1.18\,\text{ms}$, $\overline{\epsilon}_{proj}^n$ falls below the prescribed threshold (${\epsilon}_{proj} = 1 \times 10^{-6}$), which indicates that the newly acquired snapshots contribute negligible new information to enrich the candidate basis. Therefore, Stage 2 criterion is satisfied and the framework proceeds to construct static-ROM using the three different strategies introduced in Sec.~\ref{sec:SROM} with the model parameters selected for each strategy summarized in Table \ref{tab:rom_parameters}: 

\begin{table}[pos=ht]
\centering
\caption{Summary of the basis dimensions and nonlinear-model parameters used for the different ROM formulations.}
\label{tab:rom_parameters}
\begin{tabular}{p{0.1\textwidth} p{0.25\textwidth} p{0.3\textwidth} p{0.2\textwidth}}
\hline
\textbf{Method} &
\textbf{Leading linear basis} &
\textbf{Nonlinear component} & 
\textbf{Hyperparameter} \\
\hline

Linear POD &
$n_p = 44$ POD modes &
None &
None \\[2mm]

\hline
Quadratic nonlinear manifold &
$n_p = 44$ POD modes &
$\bar{n}_p = 990$ quadratic basis &
$\lambda = 1\times10^{-8}$.\\[2mm]
\hline

GPR closure &
$n_p = 11$ POD modes &
$\bar{n}_p=33$ residual modes ($n_{\mathrm{tot}}=44$)&
$\nu = 1.5$, $\ell$ = 1 \\[2mm]
\hline

RBF closure &
$n_p = 11$ POD modes &
$\bar{n}_p=33$ residual modes ($n_{\mathrm{tot}}=44$)&
$\ell = 1\times10^{-9}$ \\[2mm]
\hline

\hline
\end{tabular}
\end{table}

\begin{figure}[pos=ht] 
    \centering 
    \includegraphics[width=0.85\textwidth]{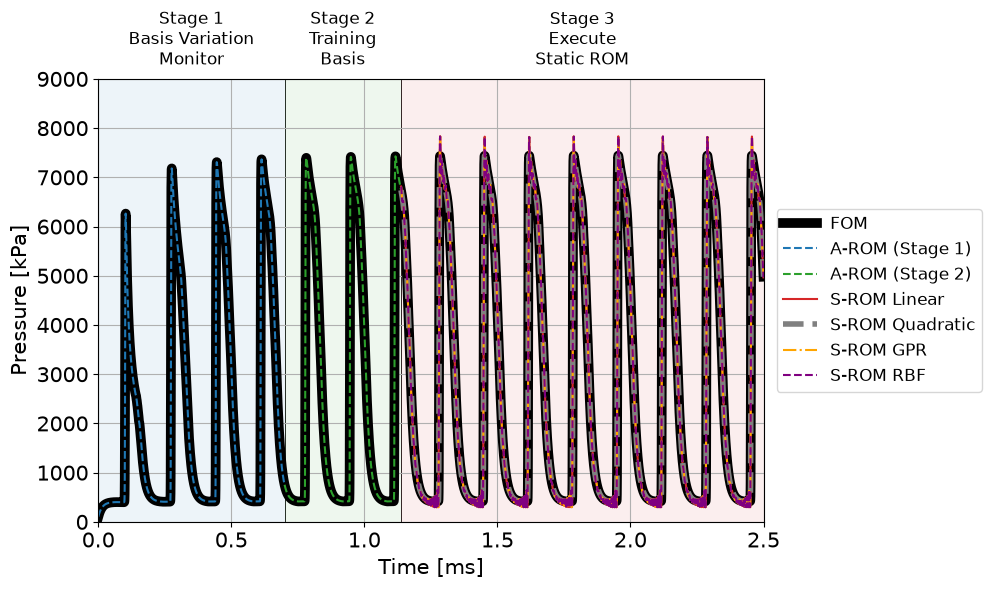} 
    \caption{Comparisons of pressure time traces with different static-ROM construction strategies for operating condition of $\phi=1$ $P_0=4 ~atm$} 
    \label{fig:trace_SROM} 
\end{figure}

\noindent With both the Stage 1 subspace-monitoring and the Stage 2 projection-error criteria satisfied, the framework transitions to \textbf{Stage 3}, where basis and sampling-point updates are deactivated and the constructed static ROM is deployed for the remainder of the simulation. Fig.~\ref{fig:trace_SROM} illustrates the resulting three-stage evolution through the pressure time trace, comparing the FOM and self-adaptive ROM solutions for different static-ROM construction strategies employed during the final stage of the framework.

\begin{figure}[pos=ht] 
    \centering 
    \includegraphics[width=0.83\textwidth]{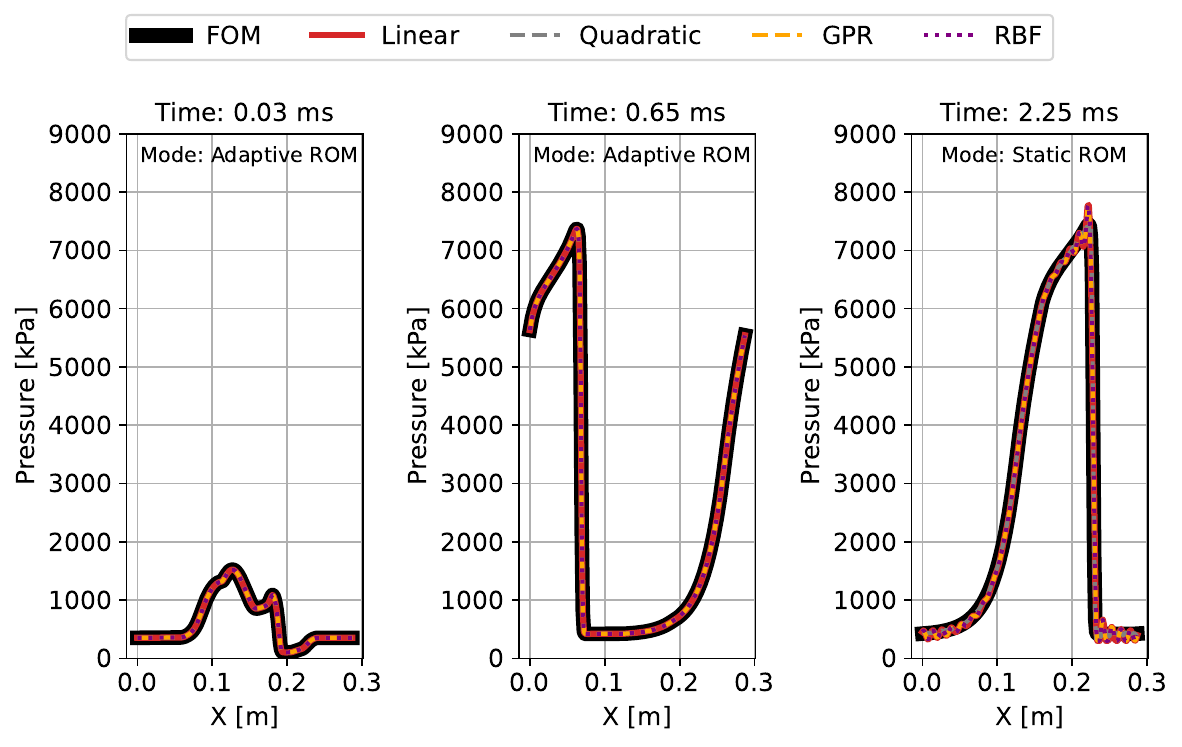} 
    \caption{Comparisons of representative FOM and self-adaptive ROM solutions for operating condition $\phi=1$ $P_0=4 ~atm$} 
    \label{fig:SROM_rep} 
\end{figure}

\begin{figure}[pos=ht] 
    \centering 
    \includegraphics[width=0.9\textwidth]{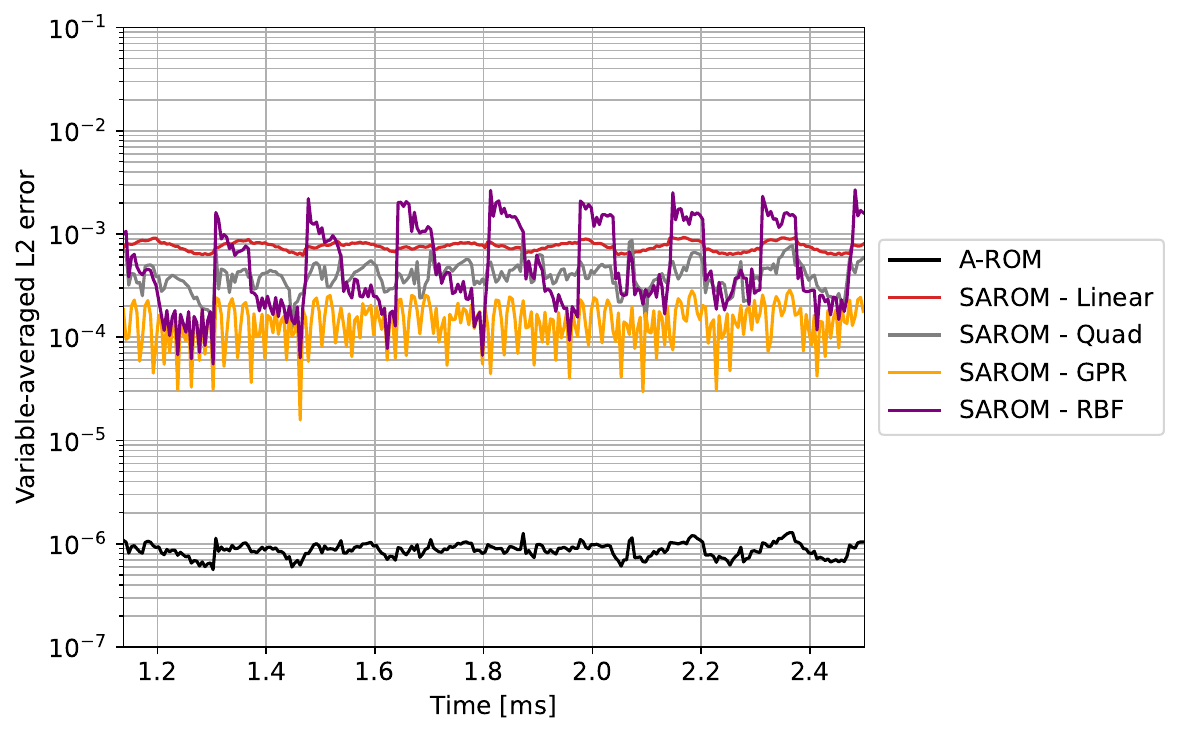} 
    \caption{Comparisons of self-adaptive ROM prediction errors with different static-ROM construction strategies for operating condition $\phi=1$ $P_0=4 ~atm$} 
    \label{fig:error_SROM} 
\end{figure}

The performance of the self-adaptive ROM is demonstrated by comparing the FOM and ROM solutions at three representative time instances in Fig.~\ref{fig:SROM_rep} corresponding to different stages of the simulation: (1) $t = 0.03\,\text{ms}$, during the detonation startup when the adaptive ROM is employed to capture the rapidly evolving dynamics; (2) $t = 0.65\,\text{ms}$, when the adaptive ROM remains active while the detonation-wave dynamics become increasingly recurrent; and (3) $t = 2.25\,\text{ms}$, after the transition to static ROM. For all three static-ROM construction strategies considered, the proposed self-adaptive ROM framework successfully completes the full $2.5\,\text{ms}$ simulation with stable and accurate predictions, while autonomously determining when the on-the-fly enriched subspace (or basis) is sufficiently representative of the encountering dynamics for transition to the static ROM. Overall, the different static-ROM strategies provide reasonably accurate predictions \emph{while} the linear-POD-basis method produces gibbs-like oscillations and overshoots near the detonation-wave front ($t = 2.25\,\text{ms}$ in Fig.~\ref{fig:SROM_rep}), a direct indicator for Kolmogorv barrier. In addition, noticeable differences can be observed in their quantitative errors (based on Eq. \ref{eq:total_err_time_avg}) as shown in Fig. \ref{fig:error_SROM}. It can been seen that the GPR-based closure method consistently yields the lowest prediction error for the 1D RDE configuration considered here. The quadratic-basis approach also generally provides improved accuracy relative to the linear POD basis; however, its substantially large nonlinear basis dimension ($\bar{n}_p = 990$ modes) introduces considerable computational and memory overhead associated with it construction and storage, making its on-the-fly construction less practical for large-scale applications. As summered in Table~\ref{tab:srom_comparison}, all the self-adaptive ROMs achieve approximately a factor 16--18 acceleration in computational time compared to FOM, which presents a significant improvement from the factor of 6 speedup when only the adaptive ROM is deployed throughout the entire simulation. This additional acceleration results from transitioning to the computationally more efficient static ROM once the recurrent system dynamics have been sufficiently captured, thereby eliminating the continued overhead associated with updating the basis and sampling points. The computational benefits of this transition are expected to become increasingly pronounced for longer-duration simulations, as the initial cost of the adaptive ROM is progressively amortized over an extended period of static-ROM execution. This characteristic makes the proposed framework particularly attractive for long-duration unsteady simulations in which recurrent dynamics persist over many characteristic cycles, such as conjugate heat transfer simulations for RDEs.

\begin{table}[pos=ht]
    \centering
    \caption{Comparison of the computational efficiency and accuracy of the static ROM formulations.}
    \label{tab:srom_comparison}
    \begin{tabular}{lccccc}
        \toprule
        & \multicolumn{4}{c}{Computational Efficiency}
        & \multicolumn{1}{c}{Accuracy} \\
        \cmidrule(lr){2-5} \cmidrule(lr){6-6}

        Method
        & \makecell{Stage 2 $\rightarrow$ 3 Transition\\Training Time \mbox{[min]}}
        & \makecell{Stage 3 Execution\\Time \mbox{[min]}}
        & \makecell{Total\\Runtime \mbox{[min]}}
        & Speedup
        & \makecell{Time-Average\\Variable-Average\\$L_2$ Error} \\

        \midrule
        FOM             & -- & -- & 504.1 & -- & -- \\
        A-ROM           & -- & -- & 73.0  &  $ 6 \times$ & $8.76 \times10^{-7}$ \\
        S-ROM Linear    & 0.116 & 3.2 & 28.8 & $18 \times$ & $7.63 \times10^{-5}$ \\
        S-ROM Quadratic & 6.966 & 14.1& 46.6 & $11 \times$ & $4.10 \times10^{-5}$ \\
        S-ROM GPR       & 0.667 & 5.1 & 31.5 & $16 \times$ & $1.49 \times10^{-5}$ \\
        S-ROM RBF       & 0.733 & 5.3 & 31.8 & $16\times$ & $6.76 \times10^{-5}$ \\
        \bottomrule
    \end{tabular}
\end{table}

\subsubsection{Evaluation of parametric predictive capabilities across various \emph{fixed} operating conditions}

\begin{figure}[pos=ht] 
   \centering 
   \includegraphics[width=\textwidth]{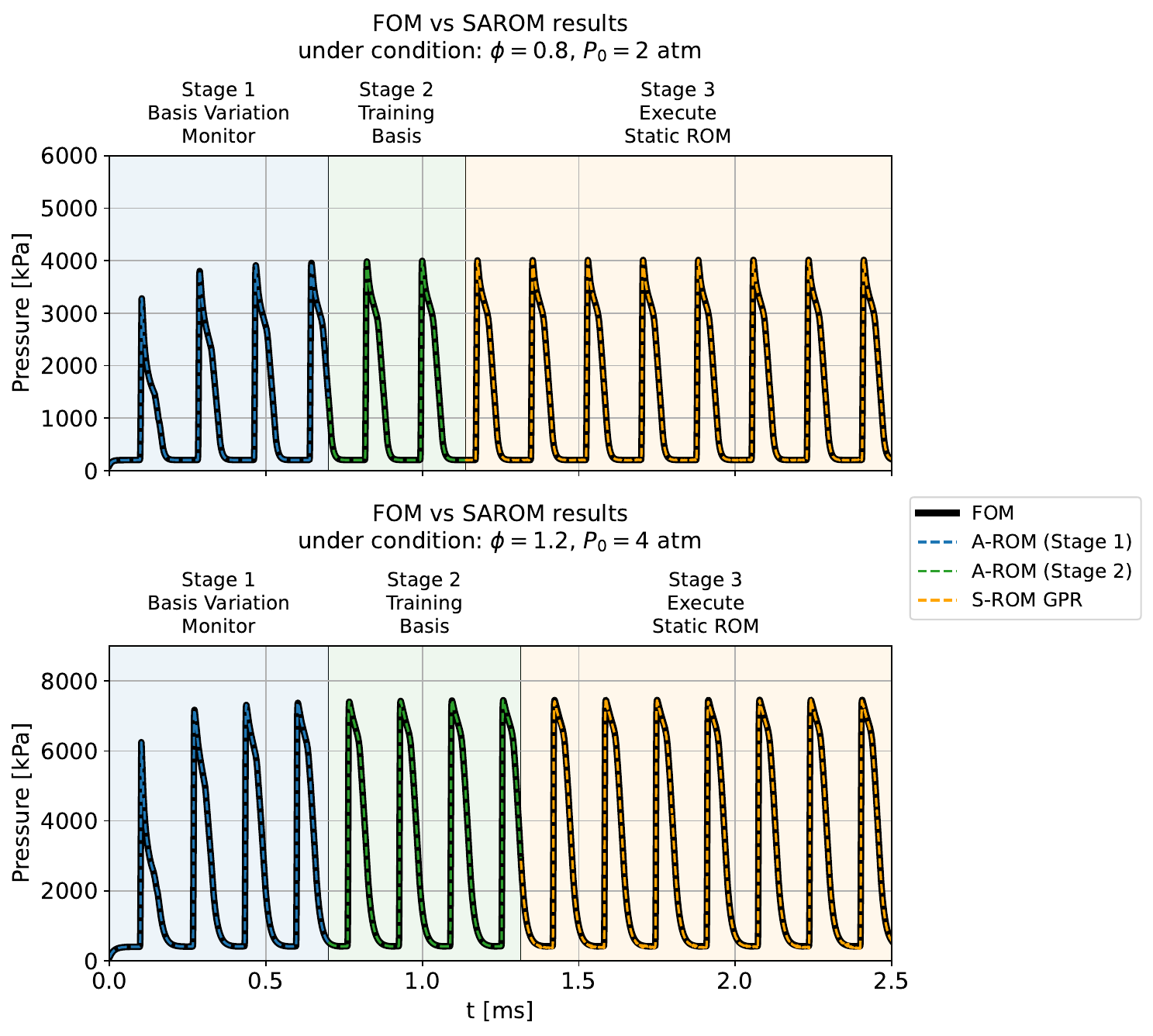} 
   \caption{Comparisons of FOM and self adaptive ROM pressure time traces under conditions: $\phi=1.2$ $P_0=4 ~atm$ and $\phi=0.8$ $P_0=2 ~atm$} 
   \label{fig:error_SROM_12_4} 
\end{figure}

\begin{figure}[pos=ht] 
   \centering 
   \includegraphics[width=0.98\textwidth]{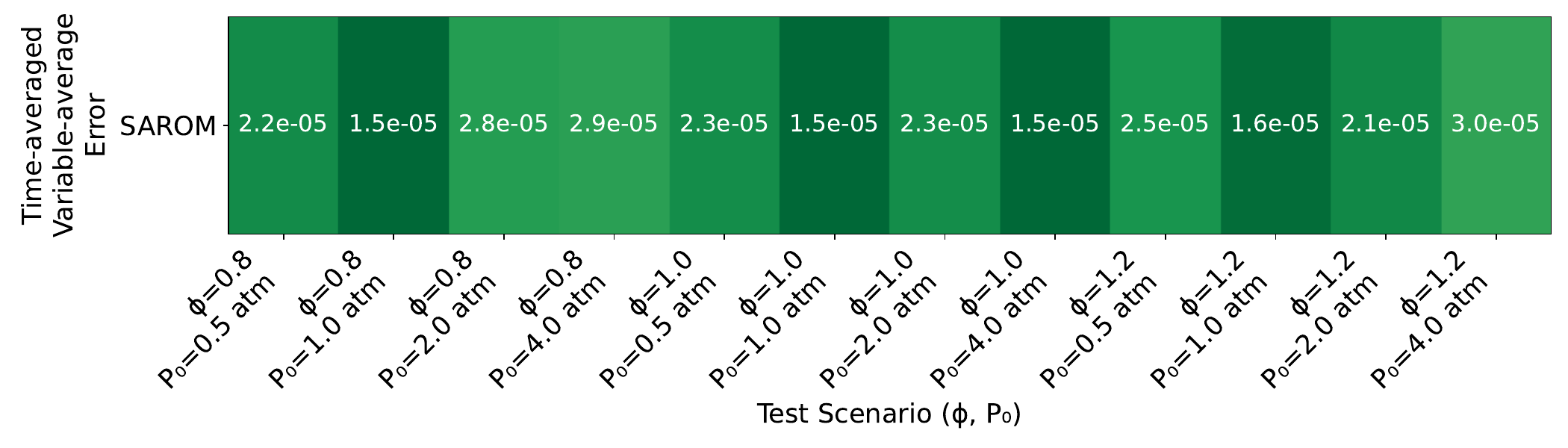} 
    \caption{Heat map of self-adaptive ROM's prediction error under all 12 operating conditions}
   \label{fig:SROM_SAROM_compare} 
\end{figure}

Next, we evaluate the predictive capabilities of self-adaptive ROM across the 12 different operating conditions introduced in Sec. \ref{sec:motiv_SAROM}. Based on the comparative assessment presented in Fig. \ref{fig:error_SROM}, GPR-based method is adopted for the static-ROM construction in the self-adaptive MOR framework. Representative pressure time traces predicted by the self-adaptive ROM are presented in Fig.~\ref{fig:error_SROM_12_4} for operating conditions $\phi = 1.2,\, P_0 = 4\,\text{atm}$ and $\phi = 0.8,\, P_0 = 2\,\text{atm}$, respectively. Despite the distinct detonation-wave dynamics associated with these operating conditions, the self-adaptive ROM accurately captures both the rapidly evolving startup transient and the subsequent establishment of statistically stationary detonation-wave dynamics. In particular, the dominant propagating flow features and their temporal evolution are accurately reproduced after the statistically stationary regime is established. Across all 12 operating conditions, the self-adaptive ROM consistently achieves an approximately 16--18 times reduction in computational time compared to FOM while maintaining prediction errors below $10^{-4}$ as summarized in Fig.~\ref{fig:SROM_SAROM_compare}, demonstrating its ability to provide both accurate and computationally efficient predictions across a broad range of operating conditions.



\subsubsection{Evaluation of predictive capabilities under \emph{transient} variation in operating conditions}
\label{sec:SAROM_trans_param}

Furthermore, we assess the modeling capabilities of self-adaptive ROM in predicting the dynamics changes due to \emph{transient} variation in operating conditions. In practical RDE operation, the injection conditions are rarely held fixed throughout the entire operation. Throttling, fuel-flow modulation and variations in plenum pressure alter the operating condition while the engine is running, which requires the detonation wave to re-adjust from one established state to another. Such scenarios are particularly challenging for conventional static ROMs because the trial basis, constructed offline from a finite set of training conditions, must represent not only the dynamics at each individual operating condition but also the transient dynamics connecting them, which are rarely included in the training dataset. 

\begin{figure}[pos=ht]
   \centering
   \includegraphics[width=0.9\textwidth]{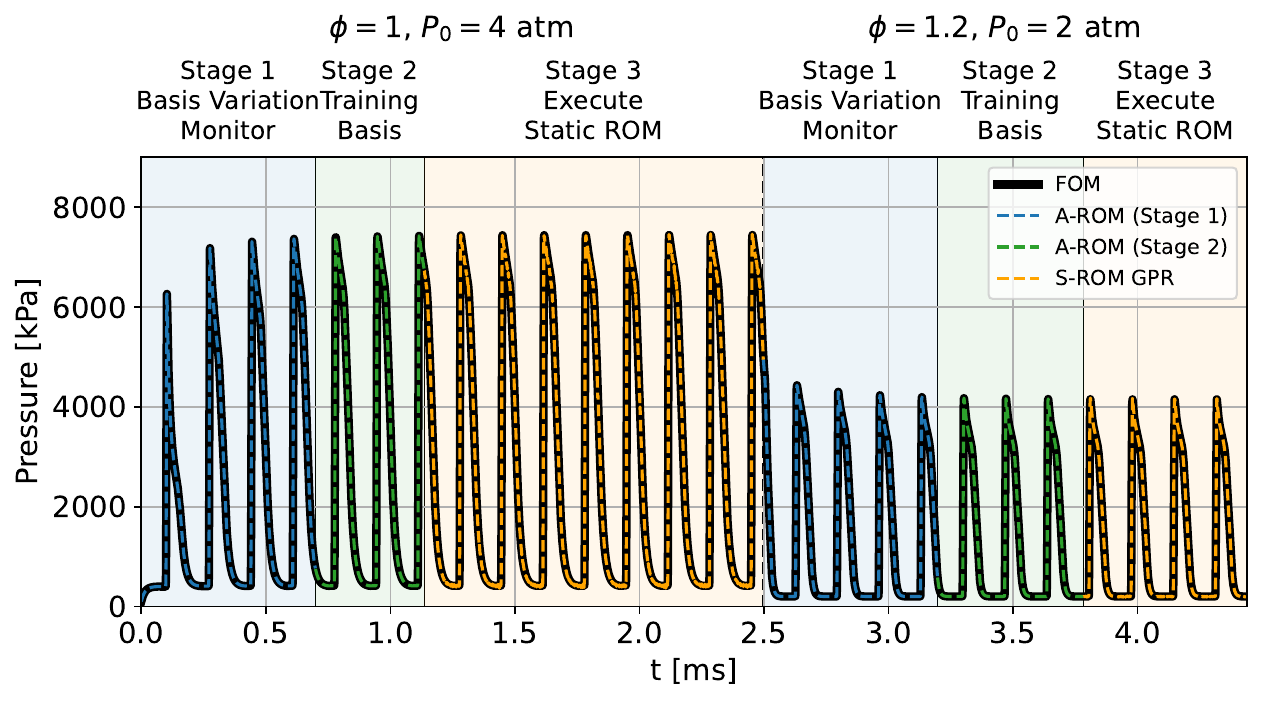}
   \caption{Comparison of FOM and self-adaptive ROM pressure time traces under a transient change in operating condition from $\phi=1$, $P_0=4~atm$ to $\phi=1.2$, $P_0=2~atm$ at $t = 2.5\,\text{ms}$.}
   \label{fig:SAROM_trans_param_var}
\end{figure}

To examine the capability of self-adaptive ROM under transient variation in operating conditions, a simulation is initialized at the baseline condition of $\phi = 1$ and $P_0 = 4\,\text{atm}$ and advanced following the same three stage procedure and model parameters described in Sec.~\ref{sec:1D_RDE}, with the GPR-based closure adopted for the static ROM construction. At $t = 2.5\,\text{ms}$, the injection parameters are abruptly changed to $\phi = 1.2$ and $P_0 = 2\,\text{atm}$, and the simulation is continued until $t = 4.5\,\text{ms}$. At $t = 2.5\,\text{ms}$, the framework deactivates the static ROM, and reverts to \textbf{Stage 1}. In this stage, the adaptive basis and sampling updates are reactivated with 10 FOM physical time steps. The candidate static basis $\mathbf{V}_{\mathrm{inc}}^{’n}$ and $\mathbf{V}_{\mathrm{inc}}^{’n}$ are reinitialized, and the principal angle monitoring between $\mathbf{V}^n$ and $\mathbf{V}^n_\mathrm{inc}$ is restarted. The same tolerances ($\epsilon_{\Phi}$, $\epsilon_{\Phi}’$, $\epsilon_{proj}$) are reused.

The resulting pressure time trace is shown in Fig.~\ref{fig:SAROM_trans_param_var}, where the three stages of the framework are indicated for both operating conditions. Before the change in injection parameters, the self-adaptive ROM behaves identically to the baseline case in Fig.~\ref{fig:trace_SROM}, transitioning from Stage 1 to Stage 2 at $t = 0.65\,\text{ms}$ and from Stage 2 to Stage 3 at $t = 1.18\,\text{ms}$. After $t = 2.5\,\text{ms}$, the detonation wave dynamics change substantially. The peak pressure decreases from approximately $7500\,\text{kPa}$ to approximately $4200\,\text{kPa}$, the post-wave pressure level also decreases, and the time interval between successive passages of the detonation wave increases, indicating a change in the detonation wave speed. These changes are consistent with the trends observed in the parametric FOM dataset (Fig.~\ref{fig:P_trace_fom_param}), where lower injection pressures lead to weaker and slower detonation waves. In addition, the first few cycles following the change exhibit a transient adjustment, in which the peak pressure gradually relaxes toward its new statistically stationary level.

As the detonation wave dynamics under the new operating condition become increasingly recurrent, the Stage 1 subspace monitoring criteria in Eq.~\ref{eq:subspace_criteria} are again satisfied at $t \approx 3.2\,\text{ms}$, which is approximately $0.7\,\text{ms}$ after the change in injection parameters and comparable to the duration of Stage 1 during the initial startup. The framework then proceeds to Stage 2, in which a new candidate static basis is constructed on the fly using incremental SVD, until the projection error criterion in Eq.~\ref{eq:proj_error_criterion} is satisfied at $t \approx 3.78\,\text{ms}$. A new GPR-based static ROM is subsequently constructed and deployed in Stage 3 for the remainder of the simulation. As shown in Fig.~\ref{fig:SAROM_trans_param_var}, the resulting static ROM accurately reproduces the FOM pressure trace under the new operating condition, including the peak pressure magnitude, the arrival time of the detonation wave and the secondary pressure shoulder behind the detonation front.

These results demonstrate that the self-adaptive MOR framework is capable of handling changes in operating conditions during the simulation, with negligible FOM training data at the new operating condition. This capability is particularly attractive for RDE applications involving throttling or other time varying injection conditions, where the relevant operating trajectories are generally not known \emph{a priori}.

\subsection{Two dimensional (2D) premixed rotating detonation engine}

\begin{figure}[pos=ht]
  \centering
    \includegraphics[width=0.8\linewidth]{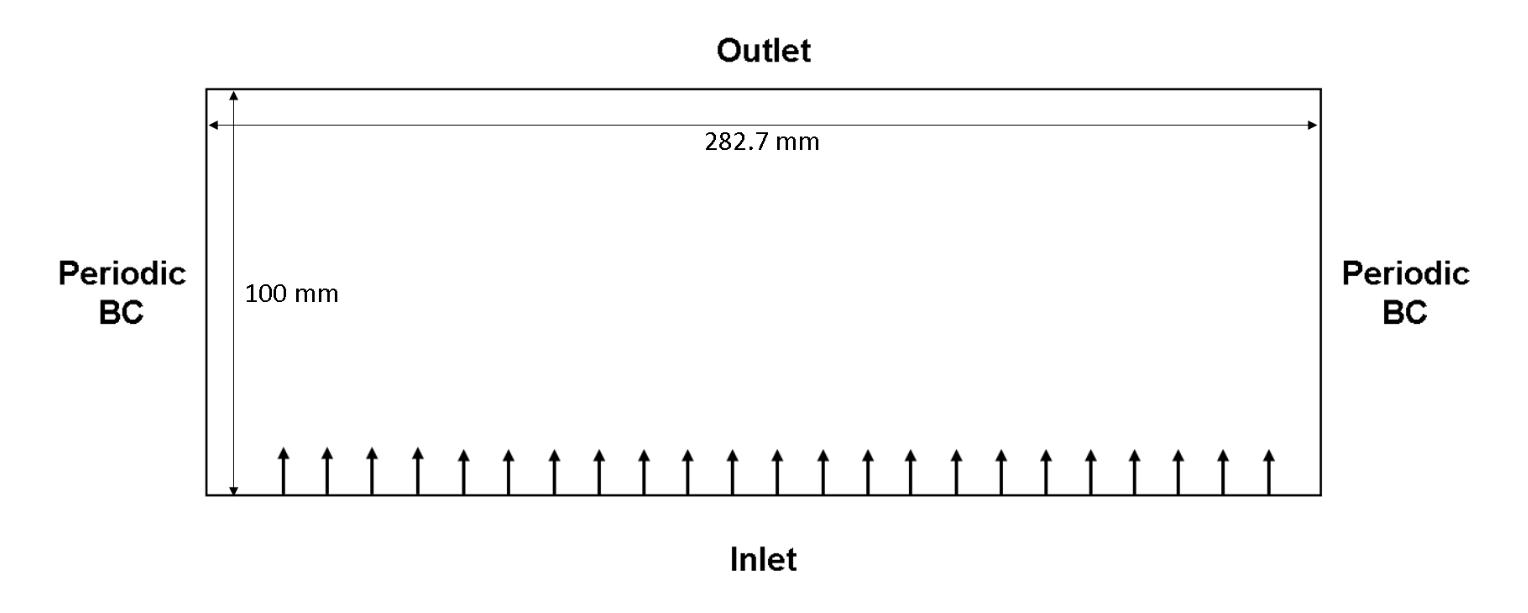}
  \caption{Computational domain of the 2D premixed RDE problem.}
 \label{fig:RDE_config}
\end{figure}

Second, we extend the demonstration of the self-adaptive ROM to a 2D premixed RDE \cite{schwer_towards_2014,schwer_numerical_2011} as shown in Fig. \ref{fig:RDE_config}, which has been previously used for demonstration of adaptive ROM techniques \cite{mohaghegh_feature-guided_2026,camacho_investigations_2024}. The computational domain extends 282.7 mm in the azimuthal direction, discretized with 1414 uniformly distributed finite-volume cells ($\Delta x = 0.2 \ \text{mm}$), and spans 100 mm in the axial direction, featuring 125 exponentially distributed cells ($\Delta y_{\text{min}} = 0.2\ \text{mm}$ and $\Delta y_{\text{max}} = 0.8 \ \text{mm}$). To account for the influence of the inlet injection, variable inflow conditions are applied at the inlet, based on injector conditions $(p, T, v)$, assuming isentropic expansion through the injector nozzle into the combustion chamber 
\begin{equation}
  \begin{cases}
    \text{if } p>p_{cr}, \text{ no flow at the inlet and boundary is treated as wall} \\ 
    \text{else }, \text{ constant velocity $V_{in}$ and temperature $T_{in}$ are applied at the inlet} 
  \end{cases}
\end{equation}

\noindent where $T_{in}$ is set to be $300$K and the critical pressure $p_{cr}$, is defined in terms of the stagnation pressure and reactant specific heat ratio, $\gamma_R$
\begin{equation}
    p_{cr} = p_0 (\frac{2}{\gamma_R+1})^{\frac{\gamma_R}{\gamma_R-1}}
\end{equation}

\noindent A fixed back pressure of 1 atm is imposed at the outlet. Combustion process is represented using a simplified single-step, two-species reaction ($\textbf{R} \rightarrow \textbf{P}$), where both the reactant (\textbf{R}) and product (\textbf{P}) are assumed to be calorically perfect gases. The thermodynamic properties of \textbf{R} and \textbf{P} are chosen to approximate a stoichiometric hydrogen-air mixture and its combustion products (nitrogen oxides and water vapor), consistent with the chemical model adopted in Schwer and Kailasanath \cite{schwer_numerical_2011}. Time integration is performed using a first-order implicit scheme with a constant time step of $\Delta t = 2\ ns$. Additional details regarding the numerical methodology and problem setup can be found in Ref \cite{camacho_investigations_2024}. The computational framework used for both FOM and self-adaptive ROM solves the conservation equations for mass, momentum, energy, and species transport in a fully coupled way using the in-house computational fluid dynamics (CFD) code, the General Mesh and Equation Solver (GEMS), which has been applied to a range of intricate, real-world reacting flow challenges \cite{harvazinski_coupling_2015,huang_combustion_2020}. All the numerical simulations are computed using 10 compute nodes (Dual Intel Xeon 6442Y CPUs) with 256GB 4800MT/s DDR5 memory.

\begin{figure}[pos=ht]
  \centering
    \includegraphics[width=\linewidth]{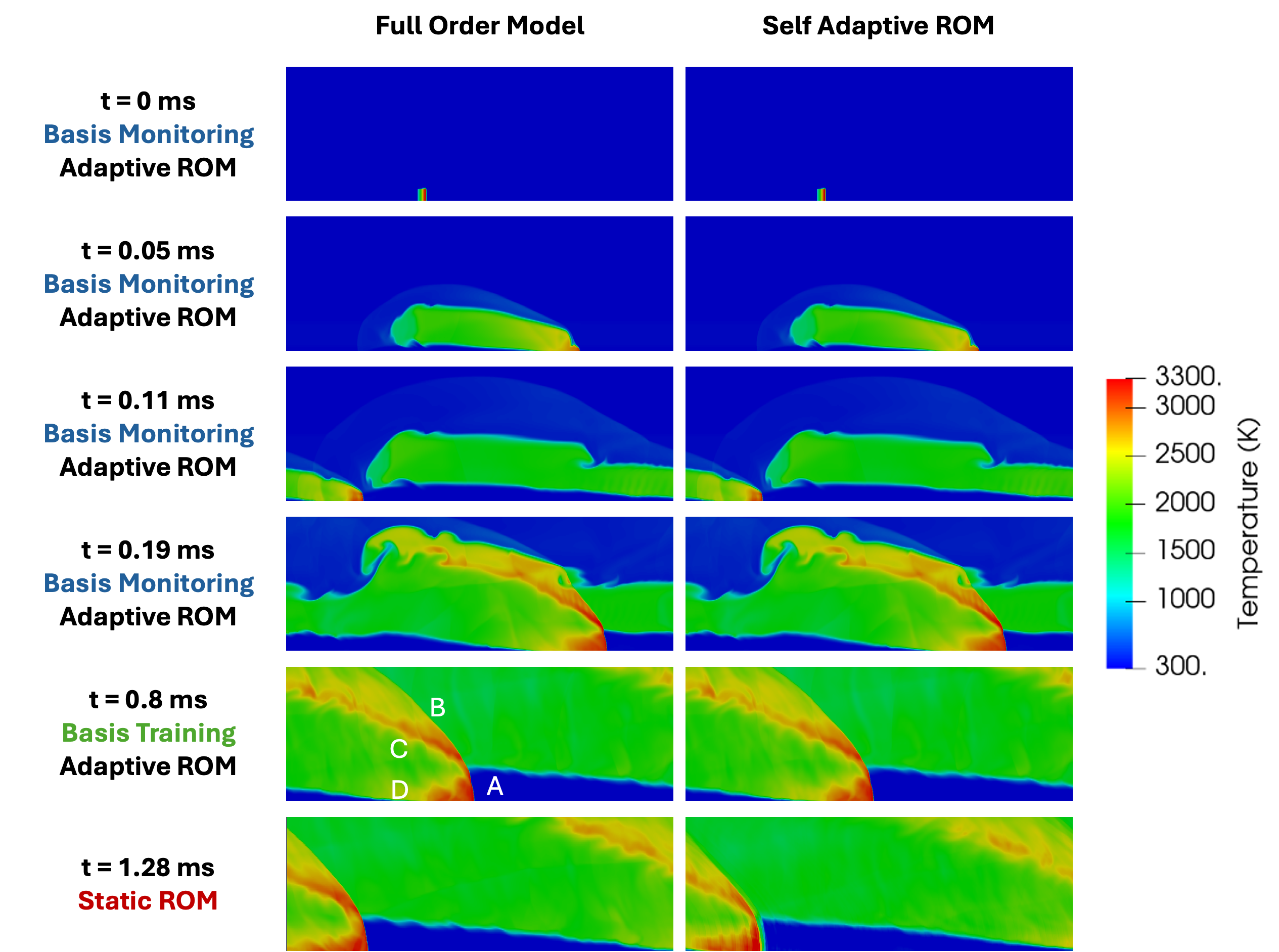}
  \caption{Comparisons of instantaneous temperature fields between FOM and self-adaptive ROM for 2D premixed RDE problem at representative time instances with
relevant features. A)detonation wave, B)oblique shock, C) shear layer, D) contact surface.}
 \label{fig:2D_RDE_temp}
\end{figure}

To demonstrate the full modeling capabilities of self-adaptive ROM for large-scale, multi-dimensional problems, we consider a 2D RDE simulation that encompasses both strongly transient and statically stationary detonation-wave dynamics as shown in the left column of Fig. \ref{fig:2D_RDE_temp}. The FOM simulation is initialized with an ignition kernel characterized by strong discontinuities with substantially elevated pressure and temperature (leading to a local pressure and temperature ratios of 75 and 10) (shown at $t = 0$ ms in Fig. \ref{fig:2D_RDE_temp} first row). An incoming velocity of 100\ $m/s$ is applied at the inlet to supply reactants to the RDE. This initialization produces a highly transient startup process involving ignition ($t = 0.05$ ms), formation and development of the detonation wave ($t = 0.11$ ms), and its subsequent transition toward a statistically stationary rotating regime ($t = 0.8$ ms). The simulation is advanced to $t = 1.7$ ms, allowing the fully developed detonation wave to complete multiple revolutions and establish sufficiently long statistically stationary state. The collection of all these dynamical behaviors in this 2D RDE problem presents a particularly challenging setting for ROM development and provides a comprehensive test case for assessing the performance of the self-adaptive ROM. The self-adaptive ROM simulation begins in the adaptive-ROM mode, initialized with 11 time step FOM calculation (i.e., $w_{init} = 11$), and constructed using 10 POD modes (i.e., $n_p = 10$) and 1767 sampling points (i.e., $n_s = 1767$, corresponding to 1\% of the entire computational domain), which are updated every 10 time steps (i.e., $z_s = 10$) with the first 10 sampling points selected using QDEIM  \cite{drmac_new_2015} and the rest determined using the FGS strategy \cite{mohaghegh_feature-guided_2026}. For the static ROM segment, a GPR-based error closure method with 189 total modes ($n_{tot}=189$) is employed. Overall, the self-adaptive ROM achieves an approximately factor of 16 speed up over FOM, 3.5 times higher than using only adaptive ROM for the entire computation. This includes 66 POD linear modes ($n_p=66$) and 123 residual modes ($\bar{n}_{p}=123$). Additionally, the linear and quadratic nonlinear manifold alternatives discussed in Sec. \ref{sec:SROM} were also implemented and evaluated. Appendix C provides a detailed comparison of these alternatives. 

\begin{figure}[pos=ht]
  \centering
    \includegraphics[width=0.95\linewidth]{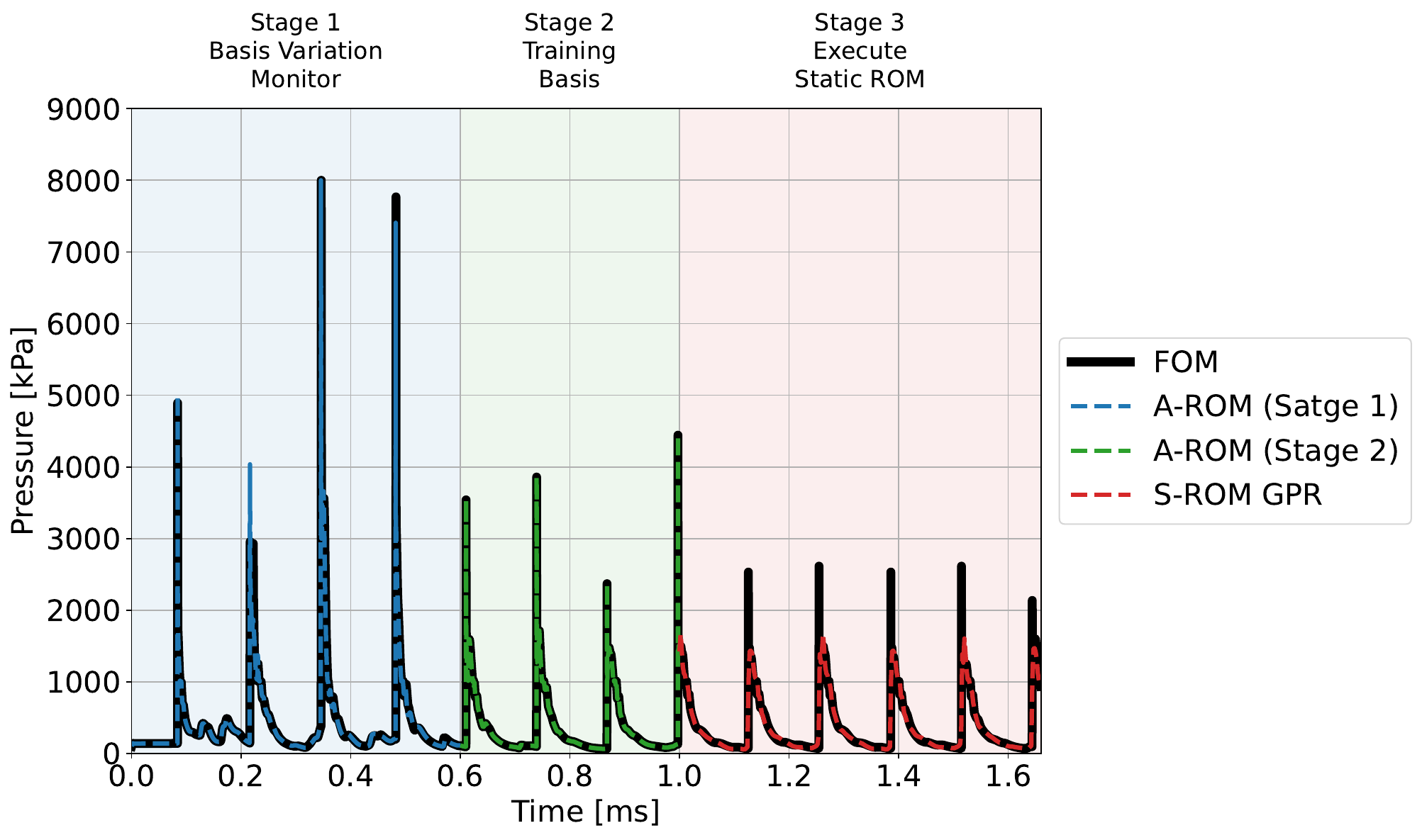}
  \caption{Comparisons of local pressure traces between FOM and self-adaptive ROM for 2D premixed RDE problem.}
 \label{fig:2D_RDE_press}
\end{figure}

Figures~\ref{fig:2D_RDE_temp} and \ref{fig:2D_RDE_press} compare the self-adaptive ROM predictions against the FOM throughout the different stages of the 2D RDE simulation. Figure~\ref{fig:2D_RDE_temp} first illustrates the evolution of the temperature field at six representative time instances spanning the initial ignition, detonation wave formation, transition toward statistically stationary dynamics, and established rotating detonation regime. The simulation begins from a quiescent domain at the fill temperature with a localized high-temperature and high-pressure ignition kernel at $t = 0$ ms. By $t=0.05$ ms, the ignition kernel produces a rapidly expanding blast wave and a region of high-temperature product gas bounded by a leading shock. At $t = 0.11$ ms, the downward propagating branch of the blast interacts with the fresh reactant layer near the injection plane and progressively develops into a self-sustained detonation. By $t = 0.19 $ ms, the flow has organized into the characteristic rotating detonation structure, consisting of a detonation front propagating through the fresh reactant layer, an oblique shock extending from the detonation front into the burned gas, and a contact surface separating different regions of combustion products. Throughout this strongly transient startup process, the adaptive ROM accurately reproduces the evolution and spatial locations of these dominant flow structures without introducing significant spurious oscillations. As the dynamics become increasingly recurrent, the framework proceeds through static basis construction and ultimately transitions to the static ROM. At $t = 0.8$ ms, during static-basis construction, and $t = 1.28$ ms, following the transition to the static ROM, the reduced-order predictions continue to reproduce the dominant detonation front, oblique shock structure, contact surface, and fresh-reactant layer observed in the FOM. Although the temperature gradients across the detonation front become slightly more diffuse during the static-ROM stage, the dominant flow structures and their relative spatial organization remain well preserved.

Figure~\ref{fig:2D_RDE_press} further quantifies these observations through the temporal pressure history recorded at a probe near the lower left corner of the computational domain, with the trace colored according to the active stage of the self-adaptive framework. During the highly transient startup phase, the signal contains large and irregularly spaced pressure peaks reaching approximately 8 MPa), nearly an order of magnitude larger than those observed after the statistically stationary regime is established. The adaptive ROM captures the timing and overall magnitude of these transient peaks reasonably well, although some peak pressures are overpredicted. As the detonation dynamics become increasingly recurrent, the pressure peaks decrease in magnitude and become more regular, with the self-adaptive ROM closely following the FOM throughout the static-basis construction stage ($0.59 \leq t \leq 1.04$ ms). Following the transition to the static ROM at ($t\approx1.04$ ms), the model continues to accurately reproduce the arrival time and temporal spacing of successive detonation wave passages, as well as the pressure evolution between successive waves, while systematically underpredicting the peak pressures. Consistent with the slightly more diffuse temperature fronts observed in Fig.~\ref{fig:2D_RDE_temp}, this reduction in peak amplitude reflects the limited ability of the truncated linear trial space to represent the sharpest gradients associated with propagating discontinuities. Importantly, the dominant characteristics of the established detonation regime: including the number of propagating waves, their propagation speed, and the corresponding cycle period, remain accurately captured. Together, these results demonstrate that the self-adaptive framework can resolve strongly evolving dynamics during startup through online adaptation and subsequently transition to an efficient static representation while preserving the essential spatial and temporal characteristics of the established detonation wave dynamics.

\section{Conclusion}
\label{sec:conclusion}

This work develops and demonstrates a self-adaptive model-order reduction (MOR) framework designed to address the longstanding trade-off between accuracy and computational efficiency in projection-based MOR, with particular emphasis on systems involving convection-dominated reacting flows such as rotating detonation engines (RDEs). Conventional adaptive reduced-order models (ROMs) can maintain high accuracy by continually updating the reduced basis tailored to the rapidly evolving local dynamics. However, frequent basis and sampling-point updates required by adaptive ROMs introduce substantial computational overhead and can diminish the computational efficiency gains obtained by ROMs. Conversely, static ROMs avoid this online adaptation cost and can provide significant computational acceleration but their accuracy may deteriorate when the solution encounters transient, multiscale, or previously unrepresented dynamics. The proposed framework reconciles these competing requirements by dynamically selecting, \emph{on the fly}, the most appropriate ROM formulation (either adaptive or static) according to the evolving characteristics of the physics encountered during the online simulation. During strongly transient regimes, the adaptive ROM can accurately resolve rapidly evolving flow structures and construct scenario-specific trial bases tailored to the local dynamics. Once the dynamics transition to recurrent or statistically stationary regimes that can be adequately represented by the constructed low-dimensional subspace, the framework switches to a static-basis ROM to reduce computational cost. This strategy eliminates the need for an extensive offline training stage while avoiding the inefficiency of continuously adapting the basis when the underlying dynamics no longer require it.

The framework's performance is evaluated using two benchmark RDE test problems,including both 1D and 2D premixed RDE configurations. For the 1D RDE problem, the framework is assessed on its capability to accommodate variations in operating conditions, including  equivalence ratios and inlet pressures. A conventional static ROM requires training data from at least 5 out of 12 FOM simulations to construct a sufficiently representative global basis and still requires approximately 180-400 POD modes to retain 99.99\% of the POD energy. In contrast, the proposed self-adaptive ROM does not require this global offline training phase, instead constructing the required low-dimensional subspace directly during the online simulation while providing very accurate predictions of parametric changes, including various fixed operating conditions and transient variation in operating conditions. The computational benefits are substantial: the self-adaptive ROM achieves approximately 18$\times$ speedup relative to the FOM, compared with approximately 6$\times$  for the conventional adaptive ROM, while maintaining comparable predictive accuracy. These computational advantages are expected to become increasingly pronounced for longer-duration simulations and problems requiring repeated evaluations across large parameter spaces, where the cost associated with continuous basis adaptation or extensive offline training becomes increasingly significant. For the 2D RDE problem, the assessment is focused on the framework's capability of modeling multi-dimensional dynamics. The self-adaptive ROM can accurately predict the entire span of the 2D detonation-wave dynamics variations, from the initial ignition, formation of detonation waves, to establishment of stationary detonation. Importantly, it achieves approximately 16$\times$ speedup relative to the FOM, compared with approximately 5$\times$  for the conventional adaptive ROM

Overall, the self-adaptive MOR framework provides a robust and scalable strategy for reduced order modeling of complex multi-physics systems by combining the complementary advantages of adaptive and static ROMs within a unified online methodology. This strategy provides a pathway toward ROM that can autonomously balance accuracy and computational efficiency without relying on extensive problem-specific offline training. Such framework is particularly promising for computationally demanding many-query applications, including parameter exploration, design optimization, uncertainty quantification, and long-time integration of large-scale systems.

\section*{Appendix A: Governing equations for full-order model}

\label{app_a}

The full-order model computations are carried out with an in-house one-dimensional CFD code developed in Python. This code solves the conservation equations for mass, momentum, energy and species mass fractions in a coupled fashion,

\[
\frac{\partial \mathbf{q}}{\partial t} + \frac{\mathbf{\partial f}}{\partial x}  - \frac{\mathbf{\partial f_{v}}}{\partial x} = \mathbf{H},
\]
where \(\mathbf{q}\) is the vector of conserved variables defined as,
\[
\mathbf{q} = \begin{pmatrix}
\rho \\
\rho u \\
\rho h^0 - p \\
\rho Y_l
\end{pmatrix},
\]
with  $\rho$  representing density, $u$ representing velocity in x-direction, \( Y_l \) representing the \( l^{th} \) species mass fraction and the total enthalpy \( h^0 \) is defined as,
\[
h^0 = h + \frac{1}{2} u^2 = \sum_{l} h_l Y_l + \frac{1}{2} u^2.
\]
The fluxes have been separated into inviscid, \(\mathbf{f}\), and viscous terms, \(\mathbf{f}_v\), The fluxes are,
\[
\mathbf{f} = \begin{pmatrix}
\rho u \\
\rho u^2 + p \\
\rho u h^0 \\
\rho u Y_l
\end{pmatrix}, \quad
\mathbf{f}_{v} = \begin{pmatrix}
0 \\
\tau_{xx} \\
u \tau_{xx} - q_x \\
\rho D_l \frac{\partial Y_l}{\partial x}
\end{pmatrix},
\]
where \(D_l\) is defined to be the diffusion of the \(l^{th}\) species into the mixture. In practice, this is an approximation used to model the multicomponent diffusion as the binary diffusion of each species into a mixture.

The heat flux in the x-direction, \(q_x\), is defined as,
\[
q_x = -K \frac{\partial T}{\partial x} + \sum_{l=1}^N \rho D_l \frac{\partial Y_l}{\partial x} h_l + q_{\text{source}},
\]
where \( K \) is the thermal conductivity, \( T \) is the temperature, \( Y_l \) are the species mass fractions, \( h_l \) are the species specific enthalpies, and \( q_{\text{source}} \) is the heat source term.
The three terms in the heat flux represent the heat transfer due to the conduction, species diffusion and heat generation from a volumetric source (e.g. heat radiation or external heat source) respectively.

The shear stress, \(\tau\), is also found in the viscous flux and defined in terms of the molecular viscosity and velocity field,

\[
\tau_{xx} = \frac{4}{3} \mu \frac{\partial u}{\partial x},
\]
where \(\mu\) is the molecular viscosity, and $u$ is the components of the velocity field.

The source term, \(\mathbf{H}\), includes a single entry for each of the species equations signifying the production or destruction of the \(l^{th}\) species, \(\dot{\omega}_l\), which is determined by the chemical kinetics,
\[
\mathbf{H} = \begin{pmatrix} 0 & 0 & 0 & \dot{\omega}_l \end{pmatrix}^T.
\]

\section*{Appendix B: Injection source-term model}
\label{app_b}

Since the one-dimensional computational domain has no physical injection boundary, reactant injection is modeled through a volumetric source term. Using an operator-splitting approach, the injection model modifies the conservative state of each injecting cell as described in this appendix. The governing equations of Appendix~A are extended to account for injection
as
\begin{equation}
  \frac{\partial \mathbf{q}}{\partial t} + \frac{\partial \mathbf{f}}{\partial x} - \frac{\mathbf{\partial f_{v}}}{x}
  = \mathbf{H} + \chi\, \mathbf{S}_{inj},
  \qquad
  \mathbf{q} = \left[\rho,\ \rho u,\ \rho E,\ \rho Y_k\right]^T ,
\end{equation}
where $\mathbf{S}_{inj}$ is the injection source term and $\chi$ is an
activation function that switches the injection on or off in each cell
according to the local pressure:
\begin{equation}
  \chi_i =
  \begin{cases}
    1, & p_i^{*} < p_{cr}, \\
    0, & p_i^{*} \ge p_{cr},
  \end{cases}
  \qquad
  p_{cr} = p_0 \left(\frac{2}{\gamma_R+1}\right)^{\frac{\gamma_R}{\gamma_R-1}} .
\end{equation}
Here, $p_0$ is the injector stagnation pressure, $\gamma_R$ is the specific
heat ratio of the reactants, and $p_i^{*}$ is the pressure of cell $i$ after
the convective and chemical sub-steps, whose state is denoted by
$\mathbf{q}_i^{*}$. The set of injecting cells is therefore
$\mathcal{I}_{inj} = \{\, i : \chi_i = 1 \,\}$. The injection step maps
$\mathbf{q}_i^{*} \rightarrow \mathbf{q}_i^{n+1}$ for all
$i \in \mathcal{I}_{inj}$ and leaves the remaining cells unchanged,
i.e., $\mathbf{q}_i^{n+1} = \mathbf{q}_i^{*}$ for $\chi_i = 0$. During one time step $\Delta t$, each injecting cell receives the mass
\begin{equation}
  \Delta m_{in} = \rho_{in}\, v_{in}\, A_{inj}\, \Delta t ,
\end{equation}
where $A_{inj}$ is the injector area associated with one cell, and
$\rho_{in}$ and $v_{in}$ are the density and velocity of the injected
reactants, obtained from isentropic expansion through the injector at
$T_{in}$. With $m_i^{*} = \rho_i^{*} V_i$, where $V_i$ is the cell
volume, the species mass fractions and specific internal energy of the
mixture follow from mass-weighted conservation:
\begin{align}
  Y_{k,i}^{mix} &= \frac{m_i^{*} Y_{k,i}^{*} + \Delta m_{in}\, Y_{k,in}}
                        {m_i^{*} + \Delta m_{in}}, \\
  e_i^{mix}     &= \frac{m_i^{*} e_i^{*} + \Delta m_{in}\, e_{in}}
                        {m_i^{*} + \Delta m_{in}},
\end{align}
where $e_{in} = e(T_{in}, p_{in}, Y_{k,in})$. The specific volume of the
mixture is
\begin{equation}
  \nu_i^{mix} = \frac{V_i}{m_i^{*} + \Delta m_{in}} .
\end{equation}
The thermodynamic state of the mixture,
$(\rho^{mix}, p^{mix}, T^{mix}, a^{mix}, \gamma^{mix})$, is then obtained
from $(e^{mix}, \nu^{mix}, Y_k^{mix})$ using the equation of state through an iterative process. Because the reactants are injected transversely, they carry no axial
momentum. Therefore, the axial velocity is
updated by assuming the stagnation pressure of the cell is preserved through
the mixing process:
\begin{equation}
  p_0^{*} = p^{*}\left(1 + \frac{\gamma^{*}-1}{2} {M^{*}}^2\right)
            ^{\frac{\gamma^{*}}{\gamma^{*}-1}},
  \qquad
  M^{n+1} = \sqrt{\frac{2}{\gamma^{n+1}-1}
     \left[\left(\frac{p_0^{*}}{p^{n+1}}\right)^{\frac{\gamma^{n+1}-1}{\gamma^{n+1}}} - 1\right]},
\end{equation}
with $u^{n+1} = M^{n+1} a^{n+1}$. The updated
state is therefore
\begin{equation}
  \mathbf{q}_i^{n+1} = \left[\rho^{mix},\ \rho^{mix} u^{*},\
      \rho^{mix}\left(e^{mix} + \tfrac{1}{2}{u^{*}}^{2}\right),\
      \rho^{mix} Y_k^{mix}\right]_i^T ,
  \qquad i \in \mathcal{I}_{inj},
\end{equation}

\section*{Appendix C: A priori offline evaluations of static manifolds' representativeness for 2D premixed RDE}
\label{app_c}

\begin{figure}[pos=ht]
  \centering
    \includegraphics[width=0.95\linewidth]{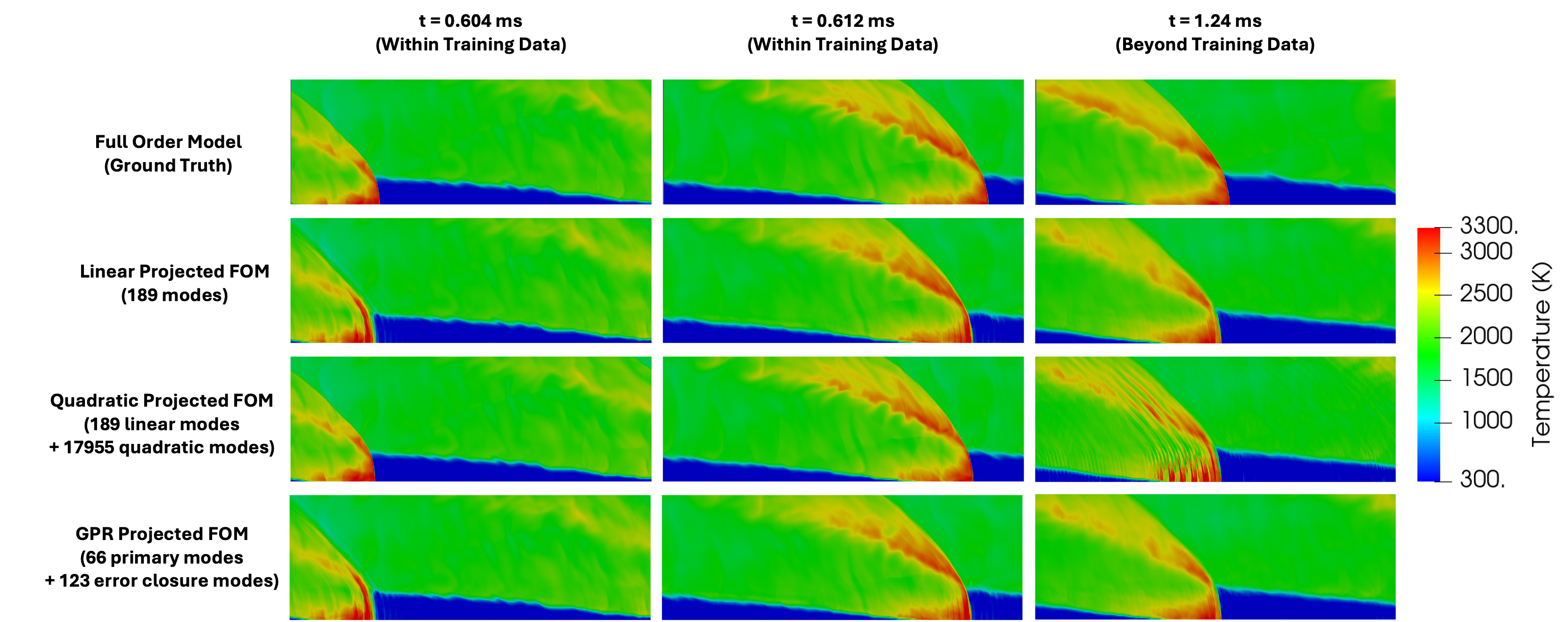}
  \caption{Comparisons of instantaneous temperature fields between the FOM and the projected FOM solutions using different basis generation strategies for the 2D RDE simulation}
 \label{fig:RDE_proj_error}
\end{figure}

For the 2D premixed RDE, an offline evaluations of the static manofilds' representativeness of detonation waves are conducted. Specifically, the FOM solution is projected onto each candidate manifold and subsequently reconstructed, thereby isolating the representation errors of the manifolds from the errors introduced by the online ROM execution. Figure \ref{fig:RDE_proj_error} compares the reconstructed fields obtained using three candidates manifolds with the FOM solution at three representative time instances, two within the training window and one beyond the training window. The GPR manifold, which augments a 66 mode linear basis with 123 error-closure modes, achieves comparable reconstruction accuracy as the 189-mode linear POD basis at all three instances, while requiring only approximately one-third of the primary latent dimension. However, it performance depends on the appropriate selection of GPR kernel and hyperparameters, and the associated tuning requirements may conflict with the goal of a self-adaptive framework that operates automatically without case-specific intervention. Similarly, the quadratic manifold provides accurate reconstruction within the training window, but its accuracy deteriorates beyond the training window, where  spurious oscillations develop near the detonation front (Fig. \ref{fig:RDE_proj_error}, bottom right of the third row). Moreover, including 189 primary linear modes in the quadratic manifold representation requires 17,955 quadratic modes, resulting in a storage requirement that scales quadratically with the number of primary modes - i.e., $O(n^2)$.

\section*{Acknowledgments}
The authors acknowledge the supports from the AFOSR under the grant FA9550-
23-1-0211 (Program managers: Drs. Chiping Li and Fariba Fahroo). All the numerical studies
were carried out under the support from the Center for Research Computing at the University of
Kansas.











\printcredits

\bibliographystyle{IEEEtran}


\bibliography{references}



\end{document}